\documentclass[prx,amsmath,amssymb, notitlepage, twocolumn,
nofootinbib,
superscriptaddress,
longbibliography
]{revtex4-1}

\usepackage{amsmath}
\usepackage{tabularx,graphicx}
\usepackage{epstopdf}
\usepackage{makecell}

\usepackage{graphicx}
\usepackage{latexsym}
\usepackage{amssymb}
\usepackage{amsmath}
\usepackage{color, colortbl}
\usepackage{mathtools}
\usepackage{psfrag}
\usepackage{bbm}
\usepackage{bm}
\usepackage{titlesec}
\usepackage{dsfont}
\usepackage{feynmp}
\usepackage{slashed}
\usepackage{multirow}
\usepackage{xurl}
\usepackage{xr}
\usepackage{xcite}

\newcommand{\onenorm}[1]{\left\| #1 \right\|_1}
\newcommand{\opnorm}[1]{\left\| #1 \right\|}

\newcommand{\beq}{\begin{eqnarray}}
	\newcommand{\eeq}{\end{eqnarray}}

\newcommand{\tr}{\mathrm {tr}}

\newcommand{\bsp}{\begin{aligned}}
	\newcommand{\esp}{\end{aligned}}

\newcommand{\ie}{{i.e., }}

\newcommand{\dist}{\mathrm{dist}}

\makeatletter
\newcommand*{\addFileDependency}[1]{
  \typeout{(#1)}
  \@addtofilelist{#1}
  \IfFileExists{#1}{}{\typeout{No file #1.}}
}
\makeatother

\newcommand*{\myexternaldocument}[1]{
    \externaldocument{#1}
    \addFileDependency{#1.tex}
    \addFileDependency{#1.aux}
}

\myexternaldocument{supplemental}

\usepackage{xcolor}
\definecolor{darkblue}{rgb}{0.,0.,0.4}
\definecolor{darkred}{rgb}{0.5,0.,0.}
\definecolor{BlueViolet}{RGB}{138,43,226}
\definecolor{SkyBlue}{RGB}{30,144,255}
\definecolor{DarkGreen}{RGB}{0,100,0}

\usepackage{xr-hyper}
\usepackage[colorlinks=true,linkcolor=darkblue,citecolor=blue,urlcolor=darkred]{hyperref}
\usepackage[normalem]{ulem}

\usepackage{mathrsfs}

\usepackage{comment}

\newcommand{\Ad}{\mathrm{Ad}}

\newtheorem{theorem}{Theorem}
\newtheorem{lemma}{Lemma}
\newtheorem{definition}{Definition}
\usepackage{pst-all}
\usepackage{leftidx}

\usepackage[off]{auto-pst-pdf} 
\usepackage{booktabs}
\usepackage{yfonts}



\myexternaldocument{Supp}

\usepackage[caption=false]{subfig}

\usepackage{enumitem}

\begin{document}

\title{Universal decay of mutual information and conditional mutual information\\
in gapped pure- and mixed-state quantum matter}

\author{Jinmin Yi}
\affiliation{Perimeter Institute for Theoretical Physics, Waterloo, Ontario, Canada N2L 2Y5}
\affiliation{Department of Physics and Astronomy, University of Waterloo, Waterloo, Ontario, Canada N2L 3G1}

\author{Kangle Li}
\affiliation{Department of Physics, National University of Singapore, 117551, Singapore}

\author{Chuan Liu}
\affiliation{Department of Physics, National University of Singapore, 117551, Singapore}

\author{Zixuan Li}
\affiliation{Department of Physics, National University of Singapore, 117551, Singapore}

\author{Liujun Zou}
\affiliation{Department of Physics, National University of Singapore, 117551, Singapore}

\begin{abstract}

For spin and fermionic systems in any spatial dimension, we establish that the superpolynomial decay behavior of mutual information and conditional mutual information is a universal property of gapped pure- and mixed-state phases, i.e., all systems in such a phase possess this property if one system in this phase possesses this property. We further demonstrate that the (conditional) mutual information indeed decays superpolynomially in a large class of phases, including chiral phases. As a byproduct, we sharpen the notion of mixed-state phases.

\end{abstract}

\maketitle

{\it Introduction} -- Entanglement has become a central tool for characterizing quantum matter, revealing nonlocal many-body properties beyond the reach of conventional order parameters. In this information theoretic framework, the mutual information (MI) and conditional mutual information (CMI) are two
fundamental measures of long-range correlations. For a tripartite state $\rho_{ABC}$ defined on regions $A,B$ and $C$, the MI between regions $A$ and $C$ is defined as
\begin{equation}
I(A:C) = S(A) + S(C) - S(AC)\;,
\end{equation} 
and the CMI is defined as 
\begin{equation}
I(A:C|B) = S(AB) + S(BC) - S(B) - S(ABC)\;,
\end{equation} 
where $S(R)$ is the von Neumann entropy of $\rho_R$, the reduced density matrix of $\rho_{ABC}$ in region $R$. The MI universally upper-bounds correlations between observables in $A$ and $C$ \cite{2008PhRvL.100g0502W}, and CMI with appropriate partitions distinguish different topological phases \cite{Kitaev2005, Levin2005}. Moreover, MI and CMI underpin the axioms of the entanglement bootstrap program~\cite{Shi2019, Shi_2021, Shi2023, yang2025topologicalmixedstatesaxiomatic}, provide robust diagnostics for mixed-state phases~\cite{Lessa_2025, sang2024stabilitymixedstatequantumphases}, and directly link to the approximate quantum error correction capabilities of many-body states~\cite{Fawzi-Renner-ACMI, Flammia_2017,Yi_2024}.

\begin{figure}
    \centering
    \includegraphics[width=0.5\linewidth]{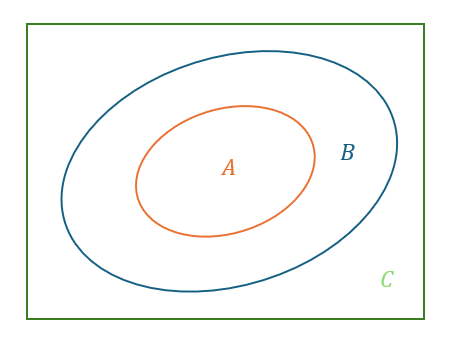}
    \caption{Region $A$ is a contractible region shielded from $C$ by the region $B$. Together, regions  $A$, $B$ and $C$ partition the entire lattice. While shown here for a 2D lattice with open boundary condition, this partition generalizes to arbitrary dimensions where the region $B$ shields a contractible $A$ from $C$ and $ABC$ together is on a closed manifold.}
    \label{fig:Def-partition}
\end{figure}

Despite their important role, the universal behaviors of MI and CMI remain largely unexplored. Consider, say, the tripartition in Fig. \ref{fig:Def-partition}. Since correlations decay exponentially in gapped phases \cite{Hastings2004}, it is often expected that MI and CMI also decay exponentially there, \ie $I\sim fe^{-d/\xi}$, with $I$ either MI or CMI, $f$ a prefactor, $d$ the distance between regions $A$ and $C$, and $\xi$ a length scale. However, small correlations do not imply small MI and CMI, and Haar-random pure states provide an illuminating example, where connected correlators are typically exponentially small, while the state has volume-law entanglement, implying large MI and CMI between macroscopic regions~\cite{Brand_o_2014}. Indeed, rigorous understanding of how MI and CMI decay is limited so far. Especially, how $f$ scales with the sizes of the subregions is poorly understood. Filling this gap in understanding is particularly urgent given that MI and CMI are now experimentally accessible (e.g., via randomized measurements, interferometric protocols, or tomography in quantum simulators~\cite{2015Nature,entropytomo_Zoller,2018Elben,2019PhRvAZoller,2020Huang,2023NatRPhPreskill,2024PhRvX,2024PRXQPreskill}), making their scaling behavior a directly testable hallmark of quantum matter.

In this work, for spin and fermionic systems in any dimension, we prove that superpolynomial decay of MI and CMI is a universal property of gapped pure- and mixed-state quantum phases (see Theorems \ref{thm:CMI_decay} and \ref{thm:mixed_CMI_decay}).  Here pure-state phases refer to ground state phases of Hamiltonians, while mixed-state phases refer to phases in open systems. The notion of ``gap" is context-dependent: For pure states it refers to the spectral gap of the Hamiltonian,{\footnote{For phases with gapless boundaries, such as chiral phases, we assume that the system is on closed manifolds so that the Hamiltonian has a positive spectral gap.}} while for mixed states it refers operationally to the presence of fast-decaying MI and CMI \cite{sang2024stabilitymixedstatequantumphases,Lessa_2025}. Our result means that if one system in such a phase has superpolynomial decaying MI or CMI, all systems in this phase share this property. Moreover, we show that a broad class of phases indeed have this property, including chiral phases where rigorous proof was previously lacking. The core idea of our proof is to note that two systems in the same phase can be connected by an adiabatic evolution, which is unitary for pure states and non-unitary for mixed states \cite{Hastings_2005,Bravyi_2010, Bachmann_2011, sang2025mixedstatephaseslocalreversibility}. We show that these adiabatic evolutions do not change MI and CMI's decay behavior substantially. Crucially, we find that the prefactor $f$ only scales polynomially with the sizes of $A$ and $B$, independent of the size of $C$ in Fig. \ref{fig:Def-partition} (see Eqs. \eqref{eq:decay1} and \eqref{eq:decay2}).

Our results are particularly timely for the study of mixed-state phases, which have received much attention recently \cite{2019Quantum_PerezGarcia,2022BarthelZhang,2022BarthelZhang2,2022QuantumSchuch,zhang2022strange,Lee_2023,Ogunnaike_2023,Fan2301,BaoFan2301,ZijianWang2024,ZijianWang2307,2023ChenGrover,2024GuWangWang,zang2024detecting,Carolyn2024,2024Sala,sang2024stabilitymixedstatequantumphases,Ma_Wang2023,2403SohalPrem,sang2024mixed_rg,Lee_2025,sala2025decoherence,Ma_Zhang_2025,guo2025locally,Ma2025,Lessa_2025,2025DaiWangWang, zhang2025fidelitystrangecorrelatorsaverage, sang2025mixedstatephaseslocalreversibility,ma2025circuitbasedchatacterizationfinitetemperaturequantum,zhang2025stabilitymixedstatephasesweak}. Previous work argued that these phases are characterized by CMI's decay behavior \cite{sang2024stabilitymixedstatequantumphases}, and proposed that two mixed states belong to the same phase if they are connected by locally reversible finite-depth quantum channels that preserve the exponential decay of CMI throughout the evolution~\cite{sang2025mixedstatephaseslocalreversibility}. Here we prove that the exponential decay behavior of MI and CMI is automatically preserved under any locally reversible finite-depth quantum channel. This finding elevates the previously conjectured universality of MI and CMI decay in mixed-state phases from an assumption to a proven theorem, and thus greatly advances our understanding of such phases.

{\it Setup and main results} -- 
In this work, we study both spin and fermionic systems on a $D$-dimensional lattice. We first focus on the ground states of almost-local Hamiltonians of the form $H=\sum_j H_j$, where the magnitude of each interaction term decays superpolynomially with the range of the interaction. Such Hamiltonians are relevant since they naturally arise in realistic settings, and the quasi-adiabatic continuation of gapped systems can be realized by finite-time evolutions generated by such Hamiltonians~\cite{Hastings_2005, Bachmann_2011, Kapustin2022Noether,hastings2010quasiadiabaticcontinuationdisorderedsystems}.

To consider the (conditional) mutual information, we focus on the partition of the lattice as in Fig. \ref{fig:Def-partition}, where $ABC$ constitutes the whole lattice that is defined on a closed manifold, and $A$ is contractible and shielded from $C$ by the region $B$. Now we state our key results on the decay behavior of MI and CMI for close quantum systems.

\begin{theorem}\label{thm:CMI_decay}
    Let $H_0$ be a gapped, almost-local Hamiltonian. Suppose that for every (possibly mixed) ground state $\rho$ of $H_0$, any of the following equations holds,
    \begin{align}\label{eq:decay1}
    I(A:C)=O(\mathrm{poly}(|A|,|B|)\dist(A,C)^{-\infty}),\\
    \label{eq:decay2}
    I(A:C| B)=O(\mathrm{poly}(|A|,|B|)\dist(A,C)^{-\infty}),
    \end{align}
    then the same equation holds for any $\rho'$ in the same gapped phase as $H_0$. Here $|A|$ and $|B|$ are the sizes of $A$ and $B$, respectively, $\dist(A, C)$ is the distance between $A$ and $C$, and $\dist(A,C)^{-\infty}$ represents functions decaying to zero faster than any polynomial.
\end{theorem}

Note that we do not restrict $\rho$ or $\rho'$ to be pure; they may be a mixed state within the ground-state subspace. If $\rho$ is pure, the MI and CMI are equivalent, \ie $I(A:C)=I(A:C|B)$.

Next, we turn to the decay behavior of MI and CMI for mixed-state phases that may not be the ground states of any Hamiltonian. The precise definition of when two mixed states are regarded as being in the same phase will be given later.

\begin{theorem}\label{thm:mixed_CMI_decay}
Let $\rho$ and $\rho'$ be two mixed states in the same phase. For the partition in Fig.~\ref{fig:Def-partition}, if $\rho$ satisfies Eq. \eqref{eq:decay1} (respectively, Eq. \eqref{eq:decay2}), then $\rho'$ also satisfies Eq. \eqref{eq:decay1} (respectively, Eq. \eqref{eq:decay2}).
\end{theorem}

{\it Decomposition of quasi-adiabatic evolution} -- Now we describe the proofs of the theorems, starting with Theorem \ref{thm:CMI_decay}. In this context, $\rho$ and $\rho'$ are related by a quasi-adiabatic continuation \cite{Hastings_2005,hastings2010quasiadiabaticcontinuationdisorderedsystems, Bachmann_2011, Kapustin2022Noether}. Our proof builds on the following intuition: MI and CMI quantify long-range correlations, which should not be generated by quasi-adiabatic evolution due to the Lieb–Robinson bound~\cite{Lieb:1972wy,Lieb-Robinson}. The cleanest case is when the quasi-adiabatic evolution is given by a finite-depth local circuit, where information propagation is confined within a lightcone. While such a circuit description alone is insufficient for our bounds for more general quasi-adiabatic evolutions, owing to large errors and uncontrollable depth, we can construct an approximate decomposition for a general quasi-adiabatic evolution that retains a sharp lightcone structure, with no leakage of information outside it.

\begin{figure*}
\centering
\includegraphics[width=\linewidth]{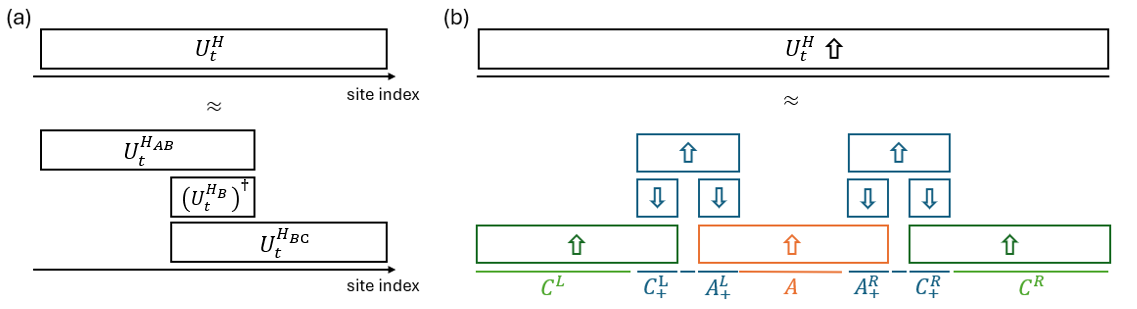}
\caption{Decomposition of the adiabatic continuation in 1D, which can be generalized to higher dimensions straightforwardly. (a) Fundamental decomposition adapted from Ref. \cite{Haah_HamiltonianEvolution}. (b) Decomposition used in this work, obtained by applying (a) four times. Regions $A$, $B$, and $C$ are shown in orange, blue, and green, respectively.}
\label{fig:LGA_decomp}
\end{figure*}

\begin{lemma}\label{lem:Utdecomp}
    For any three partitions $A,B,C$ of a lattice such that $B$ shields $A$ from $C$, define \begin{equation}
    \begin{split}
        A_+&\coloneqq\{j\in B|\dist(j,A)<\dist(A,C)/3\}\\
        C_+&\coloneqq\{j\in B|\dist(j,C)<\dist(A,C)/3\}\;.
    \end{split}
    \end{equation}

    For any evolution $U_t^H$ generated by a time-dependent almost local Hamiltonian $H$, there exists an approximate decomposition, as in Fig.~\ref{fig:LGA_decomp}(b), \ie
    \begin{equation}
        U_t^{H}\approx \tilde U_t^{\mathrm{H}}\coloneqq U_t^{H_B}\left(U_t^{H_{A_+}+H_{C_+}}\right)^\dagger U_t^{H_{CC_+}+H_{A_+A}}\;,
    \end{equation}
    where $H_R$ is the restriction of $H$ on region $R$, and the approximation error $\opnorm{U_t^H-\tilde{U}_t^H}<\epsilon$, with
    \begin{equation}
        \epsilon=O(\mathrm{poly}(|B|)\dist(A,C)^{-\infty})\;.
    \end{equation}
\end{lemma}

\begin{figure*}
\centering
\includegraphics[width=\linewidth]{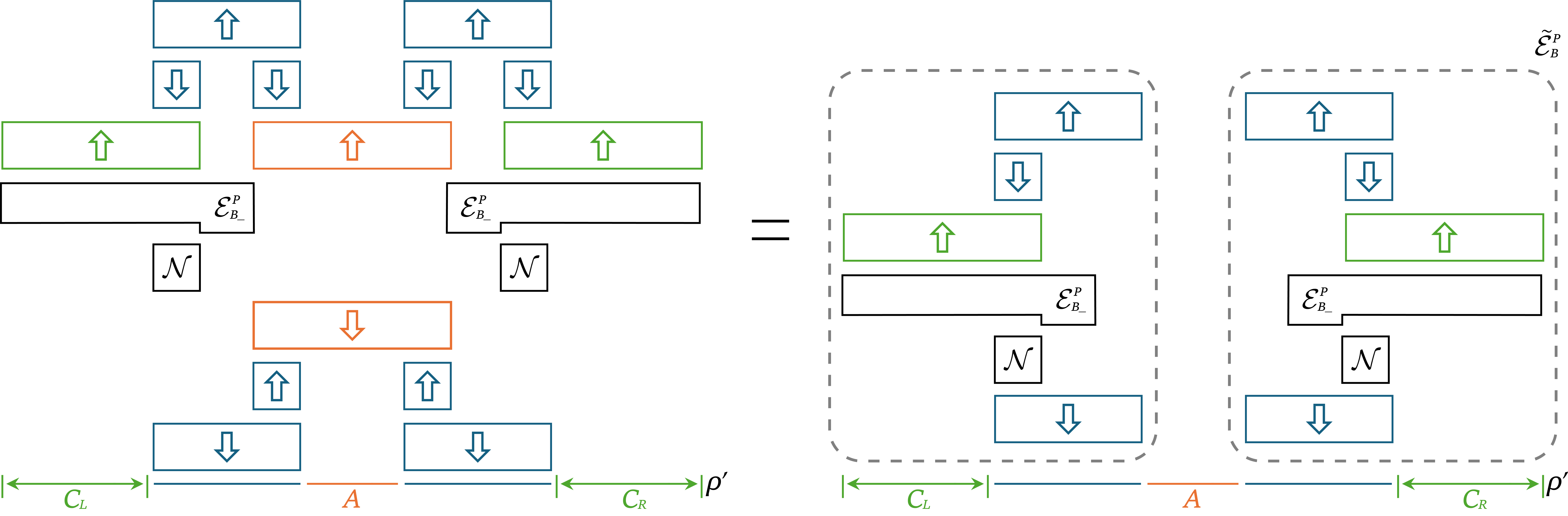}
\caption{Construction of the recovery map for local erasure noise on region $A$. Regions $A$, $B$, and $C$ are shown in orange, blue, and green, respectively. $\mathcal{N}$ denotes the erasure on region $A_+$. The Petz map $\mathcal{E}_{B_-}^P$ acts on $B_-$ with output supported on $B_-CC_+$. The idea is to evolve $\rho'$ back to $\rho$, perform the recovery, and then evolve forward to $\rho'$. On the right-hand side, certain evolutions cancel with their conjugates, yielding $\tilde{\mathcal{E}}_{B}^P$ supported entirely on region $B$.}
\label{fig:CloseSystemPetz}
\end{figure*}

The idea to prove this lemma is adapted from Ref. \cite{Haah_HamiltonianEvolution} and is largely based on the Lieb-Robinson bound for almost-local Hamiltonians~\cite{Lieb-Robinson,KapustinSopenko_Hallconductance,Kapustin2020invertible}. The key step is the decomposition illustrated in Fig.~\ref{fig:LGA_decomp}(a): the almost-local Hamiltonian evolution on a tripartite system $ABC$ is approximately factorized into a forward evolution on $BC$, a backward evolution on $B$, and then a forward evolution on $AB$. We replace the decomposition in Fig.~\ref{fig:LGA_decomp}(a) by Fig.~\ref{fig:LGA_decomp}(b), which is tailored for our proof of Theorem \ref{thm:CMI_decay}. A precise statement of this lemma, together with its proof and further discussion, is provided in the Supplementary Material (SM) \cite{supp}.

{\it Proof of Theorem 1: MI part} -- We now prove the MI part of Theorem 1. Suppose $H_0$ is a gapped, almost-local Hamiltonian. By quasi-adiabatic continuation \cite{Hastings_2005,Bravyi_2010, Bachmann_2011}, for any state $\rho'$ in the same phase as $H_0$, there exist a state $\rho$ in the ground subspace of $H_0$ and an almost-local Hamiltonian evolution $U_t^H$, such that $\rho'=U_t^H\rho (U_t^H)^\dagger$. By assumption, $\rho$ satisfies Eq. \eqref{eq:decay1}. The key idea in the proof is to find another state that satisfies Eq. \eqref{eq:decay1} and is sufficiently close to $\rho'$.

We take this state to be $\tilde{\rho}=\tilde{U}_t^H\rho(\tilde{U}_t^H)^\dagger$. To see that $\tilde\rho$ satisfies Eq. \eqref{eq:decay1}, note that under $\tilde U_t^H$ all information in $B_-\coloneqq B\backslash (A_+C_+)$ is contained in $B$ (see Fig. \ref{fig:LGA_decomp}), so $\tilde{\rho}_{AC}$ can be obtained by quantum operations acting only on $\rho_{AA_+CC_+}$. Since $I_{\tilde{\rho}}(A:C)=S(\tilde{\rho}_{AC}\|\tilde{\rho}_{A}\otimes\tilde{\rho}_{C})$, by the monotonicity of relative entropy under quantum channels \cite{Lindblad1975}, we have
\begin{equation}
    I_{\tilde{\rho}}(A:C)\leqslant I_\rho(AA_+:CC_+)
\end{equation}
Thus the MI decays superpolynomially for $\tilde{\rho}$.

On the other hand, $\opnorm{U_t^H-\tilde U_t^H}<\epsilon$ implies $\onenorm{\rho'-\tilde{\rho}}<2\epsilon$, \ie $\tilde\rho$ is close to $\rho'$. So, due to the continuity of entropy~\cite{Alicki_2004, Winter_2016}, the MI of $\rho'$ can be bounded by a function of $\epsilon$, with an additional factor of $\mathrm{poly}(|A|)$ \cite{supp}. We have thus proved the MI decay in Theorem \ref{thm:CMI_decay} for the entire phase which $H_0$ belongs to.

{\it Proof of Theorem 1: CMI part} -- We now prove the CMI part of Theorem 1. Concretely, for $\rho'=U_t^H\rho (U_t^H)^\dagger$ with $\rho$ satisfying Eq. \eqref{eq:decay2} and $U_t^H$ a finite-time evolution generated by an almost-local Hamiltonian $H$, we will show that Eq. \eqref{eq:decay2} holds for $\rho'$.

To this end, note that a small CMI is equivalent to the existence of an approximate recovery map $\mathcal{E}_B'$ for the erasure noise on $C$~\cite{Fawzi-Renner-ACMI}, \ie
\begin{equation}
\rho'\approx\mathcal{E}_B'(\rho'_{AB})\;.
\end{equation}
We claim that Eq. \eqref{eq: recover map closed} serves as such a recovery map, which is pictorially represented in Fig. \ref{fig:CloseSystemPetz}. Below we unpack this construction by explaining the intuition, while leaving the technical details to SM \cite{supp}.

For notational simplicity, denote by $\pi_R=\mathbbm{1}/d^{|R|}$ the maximally mixed state on region $R$. Note
\begin{equation}
\begin{split}
&\tr_{CC_+}\left((\tilde{U}_t^H)^\dagger(\rho'_{AB}\otimes\pi_C)\tilde{U}_t^H\right)\approx\rho_{AB\backslash C_+}\;,
\end{split}
\end{equation}
because $\tilde U^H_t\approx U^H_t$ and under $(\tilde U^H_t)^\dag$ the information in $C$ is contained in $CC_+$. Suppose $\dist(A,C)$ is large enough, Eq. \eqref{eq:decay2} for $\rho$ guarantees that there is a Petz recovery map supported on $B_-$ \cite{Fawzi-Renner-ACMI}, such that
\begin{equation}
    \rho\approx\mathcal{E}_{B_-}^P(\rho_{AB\backslash C_+})\;.
\end{equation} 
We can then use $U^H_t$ to evolve $\rho$ back to $\rho'$. We have thus constructed a recovery map for the erasure noise on region $C$ of state $\rho'$, \ie
\begin{equation}
\begin{split}
        \rho'&\approx \tilde{U}_t^H\rho(\tilde{U}_t^H)^\dagger\approx\tilde{U}_t^H\mathcal{E}_{B_-}^P(\rho_{AB\backslash C_+})(\tilde{U}_t^H)^\dagger\\
        &\approx \tilde{U}_t^H\mathcal{E}_{B_-}^P(\tr_{CC_+}\left((\tilde{U}_t^H)^\dagger(\rho'_{AB}\otimes\pi_C)\tilde{U}_t^H\right))(\tilde{U}_t^H)^\dagger\\
        &=\mathcal{E}'_B(\rho'_{AB})
\end{split}
\end{equation}
where 
\begin{equation} \label{eq: recover map closed}
    \mathcal{E}'_B\coloneqq \Ad_{U^{H_B}_t(U^{H_{C_+}}_t)^\dagger U^{H_{CC_+}}_t}\circ \mathcal{E}^P_{B_-}\circ \tr_{C_+}\circ\Ad_{(U^{H_{B}}_t)^\dagger}
\end{equation}
with $\Ad_U(\cdot)=U\cdot U^\dagger$. Crucially, $\tilde{\mathcal{E}}^P_B$ is supported on $B$ due to the cancellation of various unitaries (see Fig. \ref{fig:CloseSystemPetz}). According to Ref. \cite{Fawzi-Renner-ACMI}, the existence of this recovery map implies a small CMI for $\rho'$. In SM \cite{supp}, we show that Eq. \eqref{eq:decay2} is indeed obeyed by $\rho'$. So Theorem \ref{thm:CMI_decay} is proved.

{\it Proof of Theorem 2} --
Next, we turn to Theorem \ref{thm:mixed_CMI_decay} concerning mixed-state phases, which have acquired much interest recently. As this area is still at its initial stage, many basic definitions are still evolving. Ref. \cite{sang2025mixedstatephaseslocalreversibility} suggested a definition of mixed-state phases, which, from our perspective, should be sharpened to be the following.

\begin{definition}\label{def:mixed_reversibility}
     Two states $\rho$ and $\rho'$ are in the same phase if there exist local channel circuits $\mathcal{C}=\mathcal{C}_T \cdots \mathcal{C}_2 \mathcal{C}_1$ and $\tilde{\mathcal{C}}=\tilde{\mathcal{C}}_1\tilde{\mathcal{C}}_2\cdots \tilde{\mathcal{C}}_T $  (each $\mathcal{C}_t$ or $\tilde{\mathcal{C}}_t$ is a layer of non-overlapping local channel gates) such that:
     \begin{equation} \label{eq: two-way connection}
         \mathcal{C}(\rho)=\rho',
         \quad
         \tilde{\mathcal{C}}(\rho')=\rho.
     \end{equation}
    We also require the channels to be locally reversible, \ie for any $t$ and any $\mathcal{C}^R_t$ and $\tilde{\mathcal{C}}^R_t$ being a layer composed of a subset of gates in $\mathcal{C}_t$ and $\tilde{\mathcal{C}}_t$, respectively, with the supports of the gates fully contained in a region $R$:
    \begin{align} \label{eq: local reversibility 1}
    \tilde{\mathcal{C}}^R_t\mathcal{C}^R_t\left(\mathcal{C}_{t-1} \cdots \mathcal{C}_2 \mathcal{C}_1(\rho)\right)=\mathcal{C}_{t-1} \cdots \mathcal{C}_2 \mathcal{C}_1(\rho)\;,\\
    \label{eq: local reversibility 2}
    \mathcal{C}^R_t\tilde{\mathcal{C}}^R_t\left(\tilde{\mathcal{C}}_{t-1} \cdots \tilde{\mathcal{C}}_2 \tilde{\mathcal{C}}_1(\rho')\right)=\tilde{\mathcal{C}}_{t-1} \cdots \tilde{\mathcal{C}}_2 \tilde{\mathcal{C}}_1(\rho')\;.
    \end{align}
\end{definition}

We adopt this circuit-based definition primarily because a rigorous framework for quasi-adiabatic continuations of mixed states is currently lacking. Unlike the {\it quasi}-local dynamics relevant to pure-state phases, these finite-depth circuits with {\it strictly} local channels enforce a sharp light cone. This strict locality renders Theorem 2 still valid if the superpolynomial decay behaviors in $\dist(A, C)$ in Eqs. \eqref{eq:decay1} and \eqref{eq:decay2} are replaced by exponential or even polynomial decay behaviors.

Our definition is largely based on Ref. \cite{sang2025mixedstatephaseslocalreversibility}, but with two important differences. First, the definition in Ref. \cite{sang2025mixedstatephaseslocalreversibility}  assumed exponentially decaying CMI for states, and did not specify how the prefactor of the CMI depends on the sizes of the various regions. However, our definition applies to states with either exponentially or superpolynomially decaying CMI, with the prefactor given by Eq. \eqref{eq:decay2}.{\footnote{Technically, states in our definition can even have, for example, polynomially decaying CMI, but for such states it requires more studies to understand whether such a definition of phases is physically relevant.}}  Second, the definition in Ref. \cite{sang2025mixedstatephaseslocalreversibility} further requires that the CMI decays exponentially throughout the time evolution described by the channel circuits. Below we prove Theorem \ref{thm:mixed_CMI_decay}, which shows that this extra requirement is automatically satisfied by channels discussed in Definition \ref{def:mixed_reversibility}.

To this end, similar as before, we introduce the lightcone of regions $A$ and $C$ as $AA_+$ and $CC_+$. Concretely, suppose each gate in the local quantum channel is $k$-local, we define
\begin{equation}
\begin{split}
        A_+&=\{j\in B|\dist(j,A)\leqslant(k-1)(T-1)\},\\
        C_+&=\{j\in B|\dist(j,C)\leqslant(k-1)(T-1)\},\\
        B_-&=B\backslash{A_+C_+}.
\end{split}
\end{equation}

Eq. \eqref{eq: two-way connection} suffices to prove the MI part of Theorem \ref{thm:mixed_CMI_decay}, and Eqs. \eqref{eq: local reversibility 1} and \eqref{eq: local reversibility 2} are unnecessary. For $\rho$, $I_{\rho}(AA_+:CC_+)$ obeys \eqref{eq:decay1}, then for $\rho'=\mathcal{C}(\rho)$, the monotonicity of relative entropy implies \cite{Lindblad1975}
\begin{equation}
\begin{split}
    &I_{\rho'}(A:C)=S(\rho'_{AC}\|\rho'_A\otimes\rho_C')\\
    \leqslant&S(\mathcal{C}(\rho_{\overline{B_-}}\otimes\pi_{B_-})\|\mathcal{C}(\rho_{AA_+}\otimes\rho_{CC_+}\otimes\pi_{B_-}))\\
    \leqslant&S(\rho_{\overline{B_-}}\|\rho_{AA_+}\otimes\rho_{CC_+})=I_\rho(AA_+:CC_+)
\end{split}
\end{equation}
Thus for $\rho'$, the MI $I_{\rho'}(A:C)$ also obeys Eq. \eqref{eq:decay1}. Clearly, this still holds if the superpolynomial decay in Eq. \eqref{eq:decay1} is replaced by polynomial or exponential decay.

Turning to CMI, if $I_{\rho}(AA_+:CC_+|B_-)$ obeys Eq. \eqref{eq:decay2}, there is a Petz recovery map supported on $B_-$ that approximately recovers the erasure noise on $CC_+$ \cite{Fawzi-Renner-ACMI}. For $\rho'$, finding an approximate recovery channel for the erasure noise on $C$ enables us to prove Eq. \eqref{eq:decay2} for $\rho'$. Similar to construction for closed systems, we can first evolve $\rho'$ to $\rho$ using $\tilde{\mathcal{C}}$, then apply the Petz recovery on $\rho$, and finally return to $\rho'$ using $\mathcal{C}$. The targeted recovery channel is the composition of these operations. Eq \eqref{eq: local reversibility 1} guarantees that cancellation similar to Fig.~\ref{fig:CloseSystemPetz} occurs, so that this recovery channel is supported on $B$. Thus we can also prove Eq. \eqref{eq:decay2} for $\rho'$ \cite{supp}.

{\it States with superpolynomially decaying MI and CMI} -- Theorems \ref{thm:CMI_decay} and \ref{thm:mixed_CMI_decay} show that the decay behavior of MI and CMI in Eqs. \eqref{eq:decay1} and \eqref{eq:decay2} are universal in an entire phase. Below we establish Eqs. \eqref{eq:decay1} and \eqref{eq:decay2} for a large family of states, implying that the phases which these states belong to exhibit these properties.

For ground states of commuting-projector models describing topological orders, Eqs.~\eqref{eq:decay1} and \eqref{eq:decay2} can be verified directly. In fact, when $\dist(A,C)$ is larger than some $O(1)$ value, both MI and CMI in these models vanish, \ie $I_\rho(A:C|B)=I_\rho(A:C)=0$ \cite{Flammia_2017}. 

Some topological orders cannot be described by commuting-projector Hamiltonians, such as 2D chiral states. However, Eqs. \eqref{eq:decay1} and \eqref{eq:decay2} still hold for all bosonic 2D chiral states on closed manifolds. To see it, denote by $\rho$ such a gapped ground state and by $F$ its underlying unitary fusion category \cite{Kitaev2005}. Stacking $\rho$ with its time reversal partner $\rho_t$ leads to $\rho\otimes\rho_t$, which is a topological phase with modular tensor category $Z(F)$, the Drinfeld center of $F$ \cite{Muger2003, Davydov2010}. Since such a phase always has a representative described by the Levin-Wen commuting-projector model \cite{Kirillov2010, Balsam2010, Balsam2010a, Kirillov2011, LevinWen2005}, all states in the entire phase, including $\rho\otimes\rho_t$, satisfy Eqs. \eqref{eq:decay1} and \eqref{eq:decay2}. The von Neumann entropy in any region for $\rho\otimes\rho_t$ is twice of that for $\rho$, so $\rho$ itself satisfies Eqs. \eqref{eq:decay1} and \eqref{eq:decay2}.

More generally, it is believed (but not proved yet) that for any topological order, such as fermionic chiral phases, stacking it with its time reversal partner results in a topological phase that has a commuting-projector representative. Assuming this, all topological orders satisfy Eqs. \eqref{eq:decay1} and \eqref{eq:decay2}.

Turning to open systems, it was shown that for a large class of mixed-state phases, there exists a representative state where the MI and CMI vanish \cite{MixedStateEllison2025,yang2025topologicalmixedstatesaxiomatic}.

The above shows that Eqs. \eqref{eq:decay1} and \eqref{eq:decay2} hold for a large class of gapped pure- and mixed-state phases. However, there are gapped states violating them. Consider the Ising Hamiltonian $H=-J\sum_{\langle i,j\rangle}Z_iZ_j$, which has $(|\uparrow\uparrow\cdots\uparrow\rangle+|\downarrow\downarrow\cdots\downarrow\rangle)/\sqrt{2}$ as a ground and violates Eqs. \eqref{eq:decay1} and \eqref{eq:decay2}. But this ground state is long-range correlated and unstable against perturbations. For gapped Hamiltonians with robust ground-state subspaces, we are not aware of any counterexample to Eqs. \eqref{eq:decay1} and \eqref{eq:decay2}.

{\it Discussion} --
In this work, we show that the superpolynomial decay behavior of MI and CMI is a universal property of gapped pure- and mixed-state phases, and we verify that a large class of phases have this property. Along the way, we sharpen the definition of mixed-state phases. Our theorems can also be viewed as constraints on the amount of change of MI and CMI under time evolutions generated by either almost local Hamiltonians or locally reversible channels, which apply to states that may not be the ground states of any natural Hamiltonian.

The large-scale behavior of MI and CMI plays a central role in the entanglement bootstrap program and in the characterization of mixed-state phases \cite{Shi2019, Shi_2021, Shi2023, yang2025topologicalmixedstatesaxiomatic, Lessa_2025, sang2024stabilitymixedstatequantumphases}, for which our results provide a rigorous foundation. The decay of CMI is also closely connected to the error-correcting properties of the states we study, a perspective that has recently been proven to be fruitful in understanding quantum phases ~\cite{Yi_2024,yi2025lovaszmeetsliebschultzmattiscomplexity, Flammia_2017, sang2024approximatequantumerrorcorrecting}. Together, these connections suggest that our results can serve as a powerful guideline for classifying quantum phases through their information theoretic properties.

In the future, it is valuable to rigorously understand exactly which phases have superpolynomially decaying MI and CMI. For closed systems, standard quasi-adiabatic evolution hinges on a spectral gap, which vanishes in chiral phases with open boundaries and systems undergoing a boundary phase transition. Formulating a notion of quasi-adiabatic evolution based solely on the {\it bulk} gap is therefore crucial for understanding the decay behavior of MI and CMI on general manifolds. For open systems, we define two mixed states to be in the same phase when they are two-way connected by locally reversible finite-depth channels. In realistic settings, however, one expects that the finite-depth channels should be replaced by finite-time Lindbladian or non-Markovian evolutions. Extending the definition of mixed-state phases to incorporate these evolutions, developing a proper notion of quasi-adiabatic continuation for mixed-state phases, and clarifying the role of local reversibility and understanding the decay behavior of MI and CMI in this setting remain important future directions.

\begin{acknowledgments}
We thank Timothy H. Hsieh, Ruizhi Liu, Ruochen Ma, Bowen Shi, Chong Wang, Yifan Zhang and Yijian Zou for valuable discussions. Research at Perimeter Institute is supported in part by the Government of Canada through the Department of Innovation, Science and Economic Development and by the Province of Ontario through the Ministry of Colleges and Universities. LZ is supported in part by the National University of Singapore start-up
grants A-0009991-00-00 and A-0009991-01-00.

\end{acknowledgments}

\bibliography{lib.bib}

\end{document}


\title{Supplemental Material for ``Universal decay of mutual information and conditional mutual information in gapped pure- and mixed-state quantum matter''}
\author{Jinmin Yi}
\affiliation{Perimeter Institute for Theoretical Physics, Waterloo, Ontario, Canada N2L 2Y5}
\affiliation{Department of Physics and Astronomy, University of Waterloo, Waterloo, Ontario, Canada N2L 3G1}

\author{Kangle Li}
\affiliation{Department of Physics, National University of Singapore, 117551, Singapore}

\author{Chuan Liu}
\affiliation{Department of Physics, National University of Singapore, 117551, Singapore}

\author{Zixuan Li}
\affiliation{Department of Physics, National University of Singapore, 117551, Singapore}

\author{Liujun Zou}
\affiliation{Department of Physics, National University of Singapore, 117551, Singapore}

\maketitle
\tableofcontents\

In this supplemental material, we provide more details related to the main text. Specifically, we review the concept of almost-local operators and locally generated automorphisms (LGA) in Sec.~\ref{sec: almostlocal}. We then review the quasi-adiabatic evolution in Sec.~\ref{sec: quasi-adiabatic evolution review}. In Sec.~\ref{sec: decomposition}, we prove a decomposition theorem for LGA. In Sec.~\ref{sec:MI&CMI_projector}, we demonstrate that the MI and CMI vanish for well-separated regions in commuting-projector models that characterize topological order. In Sec.~\ref{sec: MI_decay_pure} and Sec.~\ref{sec: CMI_decay_pure}, we provide the details for the proof that the MI and CMI decay behavior is preserved under LGA, respectively. In Sec.~\ref{sec: CMI_decay_mixed}, we discuss the MI and CMI decay behavior for mixed-state phases. Finally, in Sec. \ref{secapp: fermion}, we examine that all inequalities that we use to bound the entanglement measures apply to both spin and fermionic systems.

\section{Almost local operators and locally generated automorphisms}\label{sec: almostlocal}

In many realistic setups, the individual interaction terms of a Hamiltonian are not strictly local but have tails that decay faster than any power law. Such terms are referred to as \emph{almost-local} operators~\cite{KapustinSopenko_Hallconductance}, and an \emph{almost-local Hamiltonian} is then defined as a sum of these almost-local operators that are Hermitian. Such Hamiltonians are also needed to generate quasi-adiabatic continuations. In this section, we review the key properties of almost-local operators and introduce the notion of locally generated automorphisms (LGA).

We study lattice many-body systems of bosons and fermions. For the fermionic case, since all Hamiltonians should conserve fermion parity and density matrices of fermionic systems should also have even fermion parity, we focus on operators with even fermion parity, which allows the two cases to be treated on the same footing. In the following, we will therefore not distinguish them.

Intuitively, an observable $A$ on a lattice $\Lambda$ is called almost local if it can be well approximated by a local observable. To be more precise, let us denote by $B_n(j)$ a ball of radius $n>0$ with the center at $j \in \Lambda$. That is, $B_n(j)=\{k \in \Lambda | \operatorname{dist}(k, j)<n\}$. Also, let us choose a monotonically decreasing positive (MDP) function $a(n)$ on $\mathbb{R}_{+}=[0,+\infty)$ with a superpolynomially decaying tail, \ie it is of order $O\left(n^{-\infty}\right)$ for large $n$. An observable $A$ will be called $a$-localized on a site $j$ if for any $n>0$, there is a local observable $A_n$ supported on $B_n(j)$ such that $\opnorm{A-A_n}\leqslant\opnorm{A} a(n)$. An observable will be called almost local if it is $a$-localized on $j$ for some MDP function $a(r)=O\left(r^{-\infty}\right)$ and some $j \in \Lambda$.  

Naturally, one can also represent an operator $A$ that is $a$-localized on $j$ as a sum $A \;=\;\sum_{n}A^{(n)}$, where $A^{(n)}=A_n-A_{n-1}$, with $A_n$ the above local operator that can be used to approximate $A$. Each $A^{(n)}$ is supported on the ball $B_n(j)$, with $\opnorm{A^{(n)}}< 2a(n-1)\opnorm{A}$, where we take $a(0)=1$ implicitly.

Similarly, a Hamiltonian $H=\sum_jH_j$ is called $f$-local, if the interaction term $H_j$ is $f$-localized on $j$ and uniformly bounded, \ie there exists a constant $C>0$ such that $\opnorm{H_j}\leqslant C$ for all $j$. For each almost-local interaction term, we can write $H_{j} \;=\;\sum_{n}H_{j}^{(n)}$, with $\opnorm{ H_{j}^{(n)}}<2Cf(n-1)$ and $f$ a superpolynomially decaying function. This allows us to define the restriction of a Hamiltonian $H$ on any region $R$, as 
\begin{equation}
    H_R\coloneqq\sum_j\sum_{B_n(j)\subseteq R} H_j^{(n)}
\end{equation}
Note that the choice of $H_j^{(n)}$ is not unique, and the restricted Hamiltonian $H_R$ depends on this choice. However, all we need below is the superpolynomial decay property of $H_j^{(n)}$, which holds for all choices of $H_j^{(n)}$.

A useful criterion for the almost-locality is given by the commutator bounds. The commutator of an almost-local operator and a local operator far away from it should have a superpolynomially decaying operator norm, and vice versa. Concretely, we have
\begin{lemma}[Kapustin-Sopenko~\cite{KapustinSopenko_Hallconductance}, Lemma A.1]
\label{lem:commutation=local}
Let \(A\) be an observable, \(j\in\Lambda\) a site, and \(f(r)=O(r^{-\infty})\) an MDP function.  
If for any \(k\in\Lambda\) and any \(B\in\mathcal{A}_k\) ($\mathcal{A}_k$ is the operator algebra supported at site $k$), one has
\[
\bigl\lVert [A,B]\bigr\rVert 
\;\leqslant\;2\,\lVert A\rVert\,\lVert B\rVert\,f\bigl(\mathrm{dist}(j,k)\bigr),
\]
then the observable \(A\) is \(h\)\nobreakdash-localized on site \(j\) for
\[
h(r)
=\sup_{j'\in\Lambda}\sum_{k\in\overline{B}_r(j')}f\bigl(\mathrm{dist}(j',k)\bigr)
=O(r^{-\infty}).
\]
Conversely, if \(A\) is \(f\)\nobreakdash-localized on site \(j\), then for any \(k\in\Lambda\) and any \(B\in\mathcal{A}_k\) one has
\[
\bigl\lVert [A,B]\bigr\rVert 
\;\leqslant\;2\,\lVert A\rVert\,\lVert B\rVert\,f\bigl(\mathrm{dist}(j,k)\bigr).
\]
\end{lemma}

The dynamics generated by almost-local Hamiltonians are called \emph{locally generated automorphisms} (LGA)~\cite{Kapustin2020invertible}. In this work, we are interested in almost local Hamiltonians that are time dependent and denote the time evolution generated by $H(t)$ as $U_t^H$.

Just as in local Hamiltonian evolution, we have the Lieb-Robinson bound for $U_t^H$:
\begin{lemma}[Nachtergaele et al.~\cite{Lieb-Robinson}, Theorem 2.1] \label{lem:LRbound} Take $\Lambda_1\subset\Lambda$ a finite subset of the infinite lattice. Suppose there exists a non-increasing function $F: [0,\infty)\rightarrow (0,\infty)$, such that
\begin{enumerate}
    \item $F$ is uniformly integrable in $\Lambda$, \ie
    \begin{equation}
    \|F\| \;:=\; \sup_{x \in \Lambda} \sum_{y \in \Lambda} F\bigl(d(x,y)\bigr) \;<\; \infty \,,
  \end{equation}
  \item $F$ is reproducing, \ie
  \begin{equation}
    C_F \;:=\; \sup_{x,y \in \Lambda} 
      \sum_{z \in \Lambda} 
      \frac{F\bigl(d(x,z)\bigr)\,F\bigl(d(z,y)\bigr)}{F\bigl(d(x,y)\bigr)} 
    \;<\; \infty \,.
    \label{eq:1.4}
  \end{equation}
\end{enumerate}
Then for a Hamiltonian $H=\sum_{X\subseteq\Lambda_1}H_X$ with
\begin{equation}
    \opnorm{H}_F\coloneqq \sup_{x,y\in \Lambda}\sum_{X\ni x,y}\frac{\opnorm{H_X}}{F(d(x,y))}<\infty\;,
\end{equation}
given any pair of local observables $A \in \mathcal{A}_X$ and $B \in \mathcal{A}_Y$ with $X,Y \subseteq \Lambda_1$, one has
\begin{equation}\label{eq:2.2}
\bigl\|\,[\,(U_t^H)^\dagger AU_t^H,\,B\,]\bigr\|
\;\leqslant\;
\frac{2\,\|A\|\;\|B\|}{C_F}\;g_F(t)\;\sum_{x\in X}\sum_{y\in Y}F\bigl(\dist(x,y)\bigr),
\quad
\text{for any }t \in \mathbb{R}.
\end{equation}
where the function
\begin{equation}\label{eq:2.3}
g_F(t) \;=\;
\begin{cases}
e^{\,2\|H\|_F\,C_F\,|t|}\;-\;1, 
& \text{if }\dist(X,Y) > 0,\\[1em]
e^{\,2\|H\|_F\,C_F\,|t|}, 
& \text{otherwise.}
\end{cases}
\end{equation}
\end{lemma}

One easily confirms that any almost-local Hamiltonian satisfies the conditions stated above. In later sections, whenever we apply the Lieb–Robinson bound, we will explicitly verify these conditions and compute the resulting constants $C_F$ and $\|H\|_F$.

\section{Quasi-adiabatic evolution}\label{sec: quasi-adiabatic evolution review}

Suppose two (almost) local Hamiltonians $H_0$ and $H_1$ are smoothly connected by a path $H_s$ along which the gap remains open. The ground-state subspaces of the two systems can be connected by a quasi-adiabatic evolution Hamiltonian~\cite{Hastings_2005,Bravyi_2010}. Here by smooth we assume that $\partial_s H_s$ is well defined and is also an almost-local Hamiltonian.

To construct the quasi-adiabatic evolution, we define a quasi-adiabatic continuation operator $\mathcal{D}_s$ by
\begin{equation}\label{eq:adiabatic1}
    \mathcal{D}_s \coloneqq i \int \mathrm{~d} t F(t) \exp \left(i H_s t\right)\left( \partial_sH_s\right) \exp \left(-i H_s t\right)
\end{equation}
where the function $F(t)$ is chosen with the following properties. (i) The Fourier transform of $F(t)$, denoted as $\tilde{F}(\omega)$, obeys $\tilde{F}(\omega)=-1 / \omega$ for $|\omega| \geqslant 1 / 2$; (ii) $\tilde{F}(\omega)$ is infinitely differentiable; (iii) $F(t)=-F(-t)$, so $\mathcal{D}_s$ is Hermitian. Note that properties (i) and (ii) ensure that $|F(t)|=O(|t|^{-\infty})$, since with integration by parts we have
\begin{equation}
    F(t)=\int_{-\infty}^\infty d\omega e^{-i\omega t}\tilde{F}(\omega)=\frac{1}{(it)^n}\int_{-\infty}^\infty d\omega e^{-i\omega t}\tilde{F}^{(n)}(\omega)\;.
\end{equation}
(i) and (ii) thus ensure that $|F(t)|=O(|t|^{-n})$ for any positive $n$. See Ref.~\cite{hastings2010quasiadiabaticcontinuationdisorderedsystems} for an explicit construction of $F(t)$.

We can now define the quasi-adiabatic evolution $U_s$ as follows:

\begin{equation}\label{eq:adiabatic2}
    U_s \coloneqq \mathcal{S}' \exp \left\{i \int_0^s \mathrm{~d} s^{\prime} \mathcal{D}_{s'}\right\}
\end{equation}
where $\mathcal{S}'$ denotes an $s$-ordered exponential. $U_s$ generates a continuous map between the ground subspace projectors:

\begin{lemma}[Bravyi et. al.~\cite{Bravyi_2010}, Lemma 7.1]
    Let $H_s$ be a differentiable family of Hamiltonians. Let $\left|\Psi^i(s)\right\rangle$ denote eigenstates of $H_s$ with energies $E_i$. Let $E_{\min }(s)<E_{\max }(s)$ be continuous functions of $s$ and
\begin{equation}
    I(s)=\left\{\lambda \in \mathbb{R}|E_{\min }(s) \leqslant \lambda \leqslant E_{\max }(s)\right\}
\end{equation}
Define a projector $P(s)$ onto an eigenspace of $H_s$ by
\begin{equation}
    P(s)=\sum_{i: E_i \in I(s)}\left|\Psi^i(s)\right\rangle\left\langle\Psi^i(s)\right|
\end{equation}
Assume that the space that $P(s)$ projects onto is separated from the rest of the spectrum by a gap of at least $1 / 2$ for all $s$ with $0 \leqslant s \leqslant 1$. That is, any eigenvalue of $H_s$ either belongs to $I(s)$ or is separated from $I(s)$ by a gap at least $1 / 2$. Then, for all $s$ with $0 \leqslant s \leqslant 1$, we have
\begin{equation}
    P(s)=U_s P(0) U_s^{\dagger}
\end{equation}
\end{lemma} 
In what follows, the only property of $U_s$ we need is that it is an LGA (see Lemma. F.1. in Ref. \cite{Kapustin2022Noether} for the proof).

\section{Decomposition lemma for LGA}\label{sec: decomposition}

\begin{figure}
    \centering
    \includegraphics[width=0.5\linewidth]{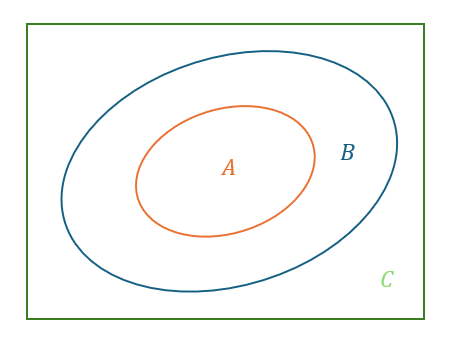}
\caption{Partition of the lattice for the definition of mutual information $I(A:C)$ and conditional mutual information $I(A:C|B)$. $A$ is a contractible region shielded from $C$ by the region $B$, and $ABC$ constitute the whole lattice. }
    \label{fig:DimCMI}
\end{figure}

In this section, we provide detailed information on the decomposition of LGA and a bound for the approximation error. 

\begin{figure}
    \centering
    \includegraphics[width=\linewidth]{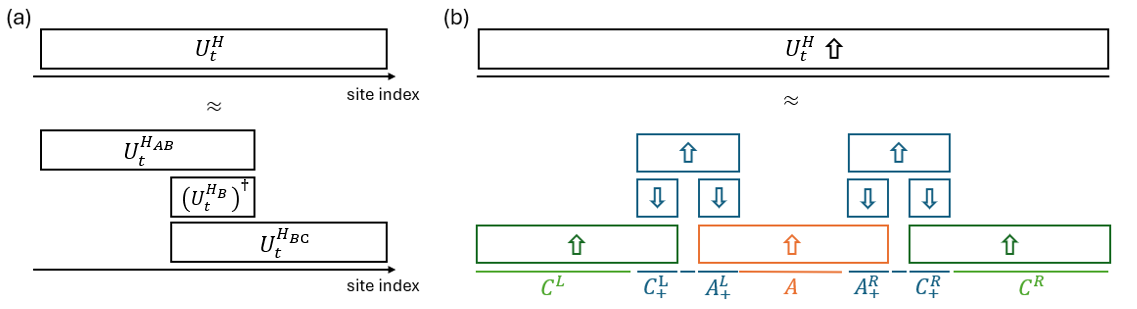}
\caption{(a) Decomposition of time evolution into three blocks, adapted from Figure 1 in Ref.~\cite{Haah_HamiltonianEvolution}. (b) Example of our LGA decomposition theorem in 1D, which can be straightforwardly generalized to any spatial dimension.}
    \label{fig:DecompositionLemma}
\end{figure}

Our proof follows the algorithm to decompose a local Hamiltonian evolution into shallow quantum circuits as described in the main text. The construction is based on a decomposition of a Hamiltonian evolution as in Fig.~\ref{fig:DecompositionLemma}(a), which was proven for strictly local Hamiltonians in  Ref.~\cite{Haah_HamiltonianEvolution}, and we will generalize it to almost-local Hamiltonians. Repeating this decomposition for four times yields the LGA decomposition:
\begin{equation}
    U_t^{H}\approx \tilde U_t^H\coloneqq U_t^{H_B}\left(U_t^{H_{A_+}+H_{C_+}}\right)^\dagger U_t^{H_{CC_+}+H_{A_+A}}\;,
\end{equation}
where
\begin{equation}
    \begin{split}
        A_+&\coloneqq\{j\in B|\dist(j,A)<\dist(A,C)/3\}\\
        C_+&\coloneqq\{j\in B|\dist(j,C)<\dist(A,C)/3\}\;.
    \end{split}
\end{equation}

For a 1D lattice system, the decomposition is depicted in Fig.~\ref{fig:DecompositionLemma}(b). 

Now we establish our decomposition theorem for LGA:

\begin{theorem}\label{Thm:LGA_decomposition}
     Consider an $f$-local time dependent Hamiltonian $H$ with interactions uniformly bounded by $C$. For any partition of the lattice to regions $A,B,C$ as in Fig.~\ref{fig:DimCMI}, the LGA $U_t^{H}$ generated by $H$ can be approximated by decomposition of Hamiltonian evolutions as in Fig.~\ref{fig:DecompositionLemma}(b) \ie
     \begin{equation}
             U_t^{H}\approx \tilde U_t^H\coloneqq U_t^{H_B}\left(U_t^{H_{A_+}+H_{C_+}}\right)^\dagger U_t^{H_{CC_+}+H_{A_+A}}
     \end{equation}
     with accuracy
\begin{equation}
     \opnorm{U_t^{H}-\tilde U_t^H}< \epsilon(w(B),|B|)\coloneqq 2|B||t|f^\delta(w(B)/3)+2|B|^2|t|F_t(w(B)/6)=O(\mathrm{poly}(|B|)\cdot|w(B)|^{-\infty})\;.
\end{equation}
Here $ f^\delta(r),F_t(r)$ are superpolynomially decaying functions that are independent of the system size,
    \begin{align}
    f^{\delta}(r)&= \sum_{n>r/2}n^DV_D2Cf(n-1)\\
        F_t(r)&=\left(|t|\sum_{d=r}^\infty V_{D} d^{D}\left[2C_\partial\tilde{f}_t(d/2)
+2\tilde{C}f^\partial(d/2)+6 \tilde{f}_t(d/2) f^\partial(d/2)\right]+2|t|C_\partial f^{\delta}(r)\right)
    \end{align}
    where $S_{D}$ denotes the area of a $D$ dimensional sphere and $V_D$ denotes the volume of a unit $D$-dimensional ball, and
    \begin{align}
        \tilde{f}_t(r)&= 2C\sup_j\sum_{k\in \bar{B}_r(j)}\left(\sum_{m=1}^{\dist(j,k)/2}\frac{f(m-1)}{C_h}(e^{2M_FC_h|t|}-1)\sum_{x\in B_m(j)}h(\dist(x,k))+\sum_{m>\dist(j,k)/2}f(m-1)\right)\\
        f^\partial(r)&=\sum_{n>r}V_D(n/2)^Df(n/2)\\
        C_\partial&=\sum_nV_D(n/2)^Df(n/2)\\
        \tilde{C}&= 2C\sum_nf(n-1)\;.
    \end{align}
    In the definition of $\tilde{f}_t(r)$,  $h(r)=cf(\frac{r-1}{2})^\alpha/r^\nu$ is another MDP function with constants $c$ , $0<\alpha<1$ and $\nu>1$, chosen such that $h(r)$ is reproducing, and
    \begin{equation}
    \begin{split}
        M_f&=\sup_{x,y\in \Lambda}\frac{2C}{h(\dist(x,y))}\sum_j\sum_{\{n:x,y\in B_n(j) \}}f(n-1)\\
        C_h &= \sup_{x,y \in \Lambda} 
      \sum_{z \in \Lambda} 
      \frac{h\bigl(d(x,z)\bigr)\,h\bigl(d(z,y)\bigr)}{h\bigl(d(x,y)\bigr)}
    \end{split}
    \end{equation}
\end{theorem}

This decomposition is an improvement from the circuit approximation of local Hamiltonian evolution~\cite{Haah_HamiltonianEvolution}, with a few modifications: (i) we allow almost-local Hamiltonian evolutions instead of only strictly local Hamiltonian evolutions; (ii) we only decompose the whole evolution into O(1) parts instead of a quantum circuit; this preserves the light cone structure that we need while the approximation error can be  still controlled.

We now establish the lemmas for the proof of this theorem.
\begin{lemma}[Haah et al.~\cite{Haah_HamiltonianEvolution}, Lemma 4]\label{lem:Haah4}
    Let $A_t$ and $B_t$ be continuous time dependent Hermitian operators, and let $U_t^A$ and $U_t^B$ with $U_0^A=U_0^B=1$ be the corresponding time evolution unitaries. Then the following hold:

(i) $W_t=\left(U_t^B\right)^{\dagger} U_t^A$ is the unique solution of $\mathbf{i} \partial_t W_t=\left(\left(U_t^B\right)^{\dagger}\left(A_t-B_t\right) U_t^B\right) W_t$ and $W_0=\mathbf{1}$.

(ii) If $\left\|A_s-B_s\right\| \leqslant \delta$ for all $s \in[0, t]$, then $\left\|U_t^A-U_t^B\right\| \leqslant t \delta$.
\end{lemma}
The proof is straightforward and can be found in Ref. \cite{Haah_HamiltonianEvolution}. The core idea is that if two Hamiltonians are close enough to each other, then their evolutions will also be close to each other. 

We will also need the following lemma resulting from the Lieb-Robinson bound (see Lemma.~\ref{lem:LRbound}):
\begin{lemma}\label{lem:LRconsequence_Ddim}
    Let $H=\sum_j H_j$ be an $f$-local time dependent Hamiltonian on a $D$-dimensional lattice. Then for any three regions $A, B, C$ with $B$ shielding $A$ from $C$, and for constant $t$, we have:
    \begin{equation}\label{eq:lieb_bound_dimD}
        \opnorm{
\left(U_t^{H_{AB}+H_C}\right)^\dagger H_{\partial}U_t^{H_{AB}+H_C}-\left(U_t^{H_{B}+H_C}\right)^\dagger H_{\partial} U_t^{H_{B}+H_C}
}\leqslant |B|^2F_t(w(B)/2)=O(|B|^{-\infty})
    \end{equation}
    where $H_\partial=H_{ABC}-H_{AB}-H_C$ denotes the Hamiltonian supported on the boundary between region $AB$ and region $C$, $w(B)=\min_{i\in A,j\in C}\dist(i,j)$ denotes the width of region $B$, and $F_t$ is defined as in Theorem.~\ref{Thm:LGA_decomposition}.
\end{lemma}
\begin{proof}[Proof of Lemma.~\ref{lem:LRconsequence_Ddim}]
Note that
\begin{equation}\label{eq:lieb_bound_target_dimD}
\begin{split}
&\opnorm{\left(U_t^{H_{AB}+H_C}\right)^\dagger H_{\partial}U_t^{H_{AB}+H_C}-\left(U_t^{H_{B}+H_C}\right)^\dagger H_{\partial} U_t^{H_{B}+H_C}}\\
=&\opnorm{\int_0^t ds\partial_s\left[\left(U_{t-s}^{H_{B}+H_C}U_s^{H_{AB}+H_C}\right)^\dagger H_{\partial}U_{t-s}^{H_{B}+H_C}U_s^{H_{AB}+H_C}\right]}\\
\leqslant&\int_0^tds\opnorm{\left(U_{t-s}^{H_{B}+H_C}U_s^{H_{AB}+H_C}\right)^\dagger\left[U_{t-s}^{H_{B}+H_C}(H_{AB}-H_B)\left(U_{t-s}^{H_{B}+H_C}\right)^\dagger ,H_{\partial} \right]\left(U_{t-s}^{H_{B}+H_C}U_s^{H_{AB}+H_C}\right)}\\
\leqslant &|t|\cdot \sup_{s\in [0,t]}\opnorm{\left[U_{t-s}^{H_{B}+H_C}(H_{AB}-H_B)\left(U_{t-s}^{H_{B}+H_C}\right)^\dagger ,H_{\partial} \right]}
\end{split}
\end{equation}
where we slightly abuse the notation to use $U_{t-s}^{H}$ to denote the time evolution from time $s$ to time $t$.

Note that $H_{AB}-H_B$ is an operator approximately localized in the region $A$, and $H_\partial$ is a sum of almost local operators near the boundary between $B$ and $C$, so the right-hand side should be upper-bounded by the Lieb-Robinson bound. In the following, we first show that $H_B+H_C$ satisfies the decaying behavior required for the Lieb-Robinson bound, and then we check the locality of $H_{AB}-H_B$ and $H_\partial$. 

We now check that $H_B+H_C$ satisfies the condition in Lemma \ref{lem:LRbound}. By definition,
\begin{equation}
        H_{B}+H_C=\sum_j\sum_{\{n|B_n(j)\subseteq B \,\mathrm{or}\,C\}}H_j^{(n)}\;,
    \end{equation}
with $\opnorm{H_j^{(n)}}< 2Cf(n-1)$. For any $f(r)$ we can choose constants $c$ , $0<\alpha<1$ and $\nu>1$ to define another MDP function $h(r)=cf(\frac{r-1}{2})^\alpha/r^\nu$ which is reproducing~\cite{hastings2010quasiadiabaticcontinuationdisorderedsystems, Kapustin2020invertible}, \ie $C_h<\infty$ is a constant independent of $n$. Since $h(r)=O(r^{-\infty})$ it is also uniformly integrable. We also have
\begin{equation}
\begin{split}
    \opnorm{H_B+H_C}_h&=\sup_{x,y\in B,C}\sum_{\substack{B_n(j)\ni x,y\\B_n(j)\subseteq B\,\mathrm{or}\,C}}\frac{\opnorm{H_j^{(n)}}}{h(\dist(x,y))}\\
    &\leqslant M_f\coloneqq\sup_{x,y\in \Lambda}\frac{2C}{h(\dist(x,y))}\sum_j\sum_{\{n:x,y\in B_n(j) \}}f(n-1)<\infty
\end{split}
\end{equation}
The above function $h$ plays the role of the function $F$ in Lemma \ref{lem:LRbound}.

Note that the sum in the definition of $M_f$ is finite because for any given $n$, the choice of $j$ such that $x,y \in B_n(j)$ is only polynomial in $n$, thus the summation converge and is a constant independent of $B,C$ or the system size $N$. The existence of a finite supreme is guaranteed by the definition of $h(r)$. More explicitly,
\begin{equation}
    \sup_{x,y\in \Lambda}\frac{2C}{h(\dist(x,y))}\sum_j\sum_{\{n:x,y\in B_n(j) \}}f(n-1)\leqslant \frac{4CV_D}{c}\sum_{n=\dist(x,y)/2}^\infty n^{D+\nu} f(n)^{1-\alpha}<\infty
\end{equation}
where $V_D$ denotes the volume of a unit $D$-dimensional ball. The factor of $2V_D n^D$ results from the choices of $j$ such that $x,y\in B_n(j)$. Thus we have shown that the evolution Hamiltonian satisfies the requirement for the Lieb-Robinson bound in Lemma \ref{lem:LRbound}.

Now we approximate $H_{AB}-H_B$ by a sum of local operators. Explicitly, we split the region $B$ into two partitions with equal widths $B_{1,2}$, formaly, we define
\begin{equation}
    B_1=\{j\in B|\dist(j,A)<w(B)/2\},\quad B_2=B\backslash B_1,
\end{equation}
and the regions $ABC$ are now split into $A,B_1,B_2,C$, with $B_1$ shielding $A$ from $B_2$ and $B_2$ shielding $B_1$ from $C$. Thus we have

\begin{equation}
    H_{AB}-H_B=\sum_j\sum_{\{n, B_n(j)\subseteq AB, B_n(j)\not\subseteq B\}}H_j^{(n)}=H_{AB_1}-H_{B_1}+\delta H_{AB}
\end{equation}
where $\delta H_{AB}\coloneqq\sum_j\sum_{\{n, B_n(j)\subseteq AB, B_n(j)\not\subseteq AB_1,B_n(j)\not\subseteq B\}}H_j^{(n)}$, and the norm of this correction term can be upper bounded, \ie
\begin{equation}
    \opnorm{\delta H_{AB}}\leqslant\sum_{n>w(B)/4}\sum_{\{j: B_n(j)\cap A\neq \varnothing,\,B_n(j)\cap B_2\neq \varnothing\}}\opnorm{H_j^{(n)}}\leqslant |B|f^{\delta}(w(B)/2)\coloneqq |B|\sum_{n>w(B)/4}(n^DV_D)2Cf(n-1)\;,
\end{equation}
where the summation is taken over $n>w(B)/4$ since we want $B_n(j)$ to overlap with both $A$ and $B_2$, and in the last inequality, we used the fact that $|B_1|<|B|$. The factor of $n^DV_D$ comes from the fact that any $j$ with $B_n(j)\cap A\neq \varnothing,\,B_n(j)\cap B_2\neq \varnothing$ must reside in $B_{n}(x)$ with some $x\in B_1$. Note that by definition $f^{\delta}(r)=O(r^{-\infty})$ is also a super-polynomial decaying function that is independent of the system size. 

Now we show that $H_\partial$ is approximately localized near the boundary between $B$ and $C$. Notice that we have the following decomposition
\begin{equation}
\begin{split}
        H_\partial&=\sum_j\sum_{\{n:B_n(j)\subseteq ABC, B_n(j)\not\subseteq AB, C\}}H_j^{(n)}\\
        &=\sum_{n}H_\partial^{(n)}\coloneqq\sum_{n}(\sum_{\substack{j\in B_{n/2}(\partial),\\B_{n-\mathrm{dist}(j,\partial)}(j)\subseteq ABC,\\B_{n-\mathrm{dist}(j,\partial)}(j)\not\subseteq AB,C}} H_j^{(n-\dist(j,\partial))})\;,
    \end{split}
    \end{equation}
    where $\partial$ denotes the boundary between $B$ and $C$. In the definition of $H^{(n)}_\partial$, the summation is taken over $j\in B_{n/2}(\partial)$ because otherwise $H_j$ would reside completely in $AB$ or $C$. 
    We first show that $H_\partial$ is a bounded operator, \ie
\begin{equation}
    \opnorm{H_\partial}\leqslant\sum_n\sum_{j\in B_{n/2}(\partial)}\opnorm{H_j^{(n- \mathrm{dist}(j,\partial))}}\leqslant |B|C_\partial\coloneqq|B|\sum_nV_D(n/2)^Df(n/2)
\end{equation}
We can also bound the tail of $H_\partial$:
\begin{equation}
    \opnorm{H_\partial-\sum_{n\leqslant r}H_\partial^{(n)}}\leqslant \sum_{n>r}\sum_{j\in B_{n/2}(\partial)}\opnorm{H_j^{(n- \mathrm{dist}(j,\partial))}}\leqslant |B|f^{\partial}(r)\coloneqq|B|\sum_{n>r}V_D(n/2)^Df(n/2)\;.
\end{equation}
Note that $f(r)=O(r^{-\infty})$, we have $f^\partial(r)=O(r^{-\infty})$, which is a superpolynomial decaying function that depends only on $f$ and $C$, but not on the regions $A$ and $B$. We also want to emphasis that while this bound is not so tight because we are bounding the size of $\partial(AB:C)$ by $|B|$, some factor of $|B|$ is still unavoidable because $H_\partial$ is not an almost local operator but a sum of almost local operators on an extended region near the boundary between $B$ and $C$.

We can then apply Lemma ~\ref{lem:LRbound} to give an upper bound for the right-hand side of (~\ref{eq:lieb_bound_target_dimD}). 
\begin{equation}
\begin{split}
    &\opnorm{\left[U_{t-s}^{H_{B}+H_C}(H_{AB}-H_B)\left(U_{t-s}^{H_{B}+H_C}\right)^\dagger ,H_{\partial} \right]}\\
    \leqslant&\sum_j\opnorm{\left[U_{t-s}^{H_{B}+H_C}\left(\sum_{\{n,B_n(j)\subseteq AB_1,B_n(j)\not\subseteq B\}}H_j^{(n)}\right)\left(U_{t-s}^{H_{B}+H_C}\right)^\dagger ,H_{\partial} \right]}
    +\opnorm{\left[U_{t-s}^{H_{B}+H_C}\delta H_{AB}\left(U_{t-s}^{H_{B}+H_C}\right)^\dagger , H_\partial\right]}\\
\leqslant&\sum_j\opnorm{\left[\tilde{H}_j ,H_{\partial} \right]}
    +2|B|^2f^{\delta}(w(B)/2)C_\partial
\end{split}
\end{equation}
where $\tilde{H}_j\coloneqq\sum_{\{n,B_n(j)\subseteq AB_1,B_n(j)\not\subseteq B\}}U_{t-s}^{H_{B}+H_C}H_j^{(n)}\left(U_{t-s}^{H_{B}+H_C}\right)^\dagger$. $\tilde{H}_j$ is a bounded operator since
\begin{equation}
    \opnorm{\tilde{H}_j}\leqslant \sum_n\opnorm{H_j^{(n)}}\leqslant \tilde{C}\coloneqq 2C\sum_nf(n-1)
\end{equation}

We now show that $\tilde{H}_j$ is an almost-local operator by the Lieb-Robinson bound. More explicitly, for any $k\in\Lambda_1$ and any local operator $\mathcal{B}\in\mathcal{A}_k$ on site $k$, by Lemma \ref{lem:LRbound}, one has
\begin{equation}
    \begin{split}
        \opnorm{\left[\tilde{H}_j,\mathcal{B}\right]}&\leqslant\sum_n\opnorm{\left[U_{t-s}^{H_{B}+H_C}H_j^{(n)}\left(U_{t-s}^{H_{B}+H_C}\right)^\dagger,\mathcal{B}\right]}\\
        &\leqslant \sum_{n=1}^{\dist(j,k)/2}\frac{2\opnorm{H_j^{(n)}}\opnorm{\mathcal{B}}}{C_h}(e^{2M_FC_h|t|}-1)\sum_{x\in B_n(j)}h(\dist(x,k))+2\sum_{n>\dist(j,k)/2}\opnorm{H_j^{(n)}}\opnorm{\mathcal{B}}\\
        &\leqslant 4C\opnorm{\mathcal{B}}\left(\sum_{n=1}^{\dist(j,k)/2}\frac{f(n-1)}{C_h}(e^{2M_FC_h|t|}-1)\sum_{x\in B_n(j)}h(\dist(x,k))+\sum_{n>\dist(j,k)/2}f(n-1)\right)
        \end{split}
\end{equation}
Applying Lemma \ref{lem:commutation=local}, we have
\begin{equation}
\begin{split}
    &\opnorm{\tilde{H}_j-\sum_{n\leqslant r}\tilde{H}_j^{(n)}}\\
    \leqslant &\tilde{f}_t(r)\coloneqq 2C\sup_j\sum_{k\in \bar{B}_r(j)}\left(\sum_{m=1}^{\dist(j,k)/2}\frac{f(m-1)}{C_h}(e^{2M_FC_h|t|}-1)\sum_{x\in B_m(j)}h(\dist(x,k))+\sum_{m>\dist(j,k)/2}f(m-1)\right) 
\end{split}
\end{equation}
Since $f(r)=O(r^{-\infty})$ is super-polynomial decaying, it is evident that $\tilde{f}_t(r)=O(r^{-\infty})$ is also super-polynomial decaying. 

Thus we have
\begin{equation}
\begin{split}
    &\opnorm{\left[\tilde{H}_j ,H_{\partial} \right]}=\opnorm{\left[\tilde{H}_j-\sum_{n\leqslant \dist(j,\partial)/2}\tilde{H}_j^{(n)}+ \sum_{n\leqslant \dist(j,\partial)/2}\tilde{H}_j^{(n)},H_{\partial} -\sum_{n\leqslant \dist(j,\partial)/2}H_\partial^{(n)}+ \sum_{n\leqslant \dist(j,\partial)/2}H_\partial^{(n)}\right]}\\
    \leqslant
    &2\opnorm{\tilde{H}_j-\sum_{n\leqslant \dist(j,\partial)/2}\tilde{H}_j^{(n)}}\opnorm{H_{\partial} -\sum_{n\leqslant \dist(j,\partial)/2}H_\partial^{(n)}}+2\opnorm{\sum_{n\leqslant \dist(j,\partial)/2}\tilde{H}_j^{(n)}}\opnorm{H_{\partial} -\sum_{n\leqslant \dist(j,\partial)/2}H_\partial^{(n)}}\\
    +&2\opnorm{\tilde{H}_j-\sum_{n\leqslant \dist(j,\partial)/2}\tilde{H}_j^{(n)}}\opnorm{\sum_{n\leqslant \dist(j,\partial)/2}H_\partial^{(n)}}\\
    \leqslant &|B|\left[2\tilde{f}_t(\dist(j,\partial)/2) f^\partial(\dist(j,\partial)/2)+2(\tilde{C}+\tilde{f}_t(\dist(j,\partial)/2))f^\partial(\dist(j,\partial)/2)+ 2(C_\partial+f^\partial(\dist(j,\partial)/2))\tilde{f}_t(\dist(j,\partial)/2)\right]\\ 
    =&|B|\left[2C_\partial\tilde{f}_t(\dist(j,\partial)/2)
+2\tilde{C}f^\partial(\dist(j,\partial)/2)+6 \tilde{f}_t(\dist(j,\partial)/2) f^\partial(\dist(j,\partial)/2)  \right]
\end{split}
\end{equation}
by decomposing both operators to their restriction to two sums of local operators with non-overlapping support and the correction part.

We are finally ready to derive the bound in Eq. (\ref{eq:lieb_bound_dimD}), since
\begin{equation}
\begin{split}
    &\opnorm{
\left(U_t^{H_{AB}+H_C}\right)^\dagger H_{\partial}U_t^{H_{AB}+H_C}-\left(U_{t}^{H_{B}+H_C}\right)^\dagger H_{\partial} U_{t}^{H_{B}+H_C}
}\leqslant |t|\sum_j\opnorm{\left[\tilde{H}_j ,H_{\partial} \right]}
        +2|t||B|^2f^{\delta}(w(B)/2)C_\partial\\
        \leqslant&|B|^2F_t(w(B)/2)\coloneqq|B|^2\left(|t|\sum_{d=w(B)/2}^\infty V_{D} d^{D}\left[2C_\partial\tilde{f}_t(d/2)
+2\tilde{C}f^\partial(d/2)+6 \tilde{f}_t(d/2) f^\partial(d/2)\right]+2|t|C_\partial f^{\delta}(w(B)/2)\right)
\end{split}
\end{equation}
where $S_{D}$ denotes the area of a $D$ dimensional sphere. We have also used the fact that the number of sites which are $d$-close to the boundary $\partial(AB:C)$ is at most $V_{D} d^D|\partial(AB:C)|\leqslant V_{D} d^{D}|B|$.
\end{proof}

We now move on to our final lemma, which corresponds to the decomposition as in Fig.~\ref{fig:DecompositionLemma}(a):

\begin{lemma}\label{lem:decomposition_Ddim}
     Let $H=\sum_j H_j$ be an $f$-local time dependent Hamiltonian on a $D$-dimensional lattice. Then for any three regions $A, B, C$ with $B$ shielding $A$ from $C$, and for constant $t$, we have

\begin{equation}
\left\|U_t^{H_{AB}}\left(U_t^{H_B}\right)^{\dagger} U_t^{H_{BC}}-U_t^{H_{ABC}}\right\| \leqslant  |B||t|f^\delta(w(B))+|B|^2|t|F_t(w(B)/2)=O(|B|^{-\infty})
\end{equation}
where $f^\delta$ and $F_t$ are defined as in Lemma \ref{lem:LRconsequence_Ddim}.
\end{lemma}
\begin{proof}
    Defining $W_t\coloneqq(U_t^{H_{AB}+H_C})^\dagger U_t^{H_{ABC}}$, it follows that $ U_t^{H_{ABC}}=U_t^{H_{AB}+H_C}W_t$. By Lemma \ref{lem:Haah4} (i), $W_t$ is the unique solution of
    \begin{equation}
        \mathbf{i} \partial_t W_t=\left(\left(U_t^{H_{AB}+H_C}\right)^{\dagger}H_{\partial} U_t^{H_{AB}+H_C}\right) W_t\;,
    \end{equation}
    where $H_\partial=H_{ABC}-H_{AB}-H_C$. By Lemma \ref{lem:LRconsequence_Ddim}, we have
    \begin{equation}
        \opnorm{
\left(U_t^{H_{AB}+H_C}\right)^\dagger H_{\partial}U_t^{H_{AB}+H_C}-\left(U_t^{H_{B}+H_C}\right)^\dagger H_{\partial} U_t^{H_{B}+H_C}
}\leq |B|^2|t|F_t(w(B)/2)
    \end{equation}
    Note that $W_t$ is the unitary evolution generated by the first term, so we now consider the unitary generated by the second term. By Lemma.~\ref{lem:Haah4}(i), the unitary generated is 
    \begin{equation}
        \left(U_t^{H_{B}+H_C}\right)^\dagger U_t^{H_{B}+H_C+H_\partial}=\left(U_t^{H_{B}+H_C}\right)^\dagger U_t^{H_{ABC}-H_{AB}+H_B}
    \end{equation}
    Note that if $H$ is strictly local then we have $H_{ABC}-H_{AB}+H_B=H_{BC}$. For $f$-local Hamiltonian, we can define
    \begin{equation}
        \delta H_{BC}\coloneqq H_{ABC}-H_{AB}+H_B-H_{BC}=\sum_{j\in ABC}\sum_{\{n:B_n(j)\cap A,B,C\neq \varnothing\}}H_j^{(n)}\;.
    \end{equation}
    This correction Hamiltonian is a bounded operator, \ie
    \begin{equation}
        \opnorm{\delta H_{BC}}\leqslant \sum_{n>w(B)/2}\sum_{j\in ABC, B_n(j)\cap A,B,C\neq \varnothing}\opnorm{H_j^{(n)}}\leqslant\sum_{n>w(B)/2} |B|V_Dn^D2Cf(n-1)=|B|f^\delta(w(B))
    \end{equation}
    By Lemma \ref{lem:Haah4}(ii), we have
    \begin{equation}
        \opnorm{\left(U_t^{H_{B}+H_C}\right)^\dagger U_t^{H_{B}+H_C+H_\partial}-\left(U_t^{H_{B}+H_C}\right)^\dagger U_t^{H_{BC}}}=\opnorm{U_t^{H_{BC}+\delta H_{BC}}-U_t^{H_{BC}}}\leqslant |B||t|f^\delta(w(B))\;.
    \end{equation}
    By the same lemma, we also have
    \begin{equation}
    \begin{split}
        &\opnorm{W_t-\left(U_t^{H_{B}+H_C}\right)^\dagger U_t^{H_{B}+H_C+H_\partial}}\\
        \leqslant& |t|\opnorm{
    \left(U_t^{H_{AB}+H_C}\right)^\dagger H_{\partial}U_t^{H_{AB}+H_C}-\left(U_t^{H_{B}+H_C}\right)^\dagger H_{\partial} U_t^{H_{B}+H_C}
}
\leqslant |B|^2|t|F_t(w(B)/2)
    \end{split}
    \end{equation}
    By the triangle in equality, we thus have
    \begin{equation}
        \opnorm{W_t-\left(U_t^{H_{B}+H_C}\right)^\dagger U_t^{H_{BC}}}\leqslant |B||t|f^\delta(w(B))+|B|^2|t|F_t(w(B)/2)\;.
    \end{equation}
    On the other hand, the left hand side is
    \begin{equation}
    \begin{split}
        \opnorm{W_t-\left(U_t^{H_{B}+H_C}\right)^\dagger U_t^{H_{BC}}}&=\opnorm{\left(U_t^{H_{AB}+H_C}\right)^\dagger U_t^{H_{ABC}}-\left(U_t^{H_{B}+H_C}\right)^\dagger U_t^{H_{BC}}}\\
        &=\opnorm{ U_t^{H_{ABC}}-U_t^{H_{AB}+H_C}\left(U_t^{H_{B}+H_C}\right)^\dagger U_t^{H_{BC}}}\\
        &=\opnorm{ U_t^{H_{ABC}}-U_t^{H_{AB}}U_t^{H_C}\left(U_t^{H_{B}}U_t^{H_C}\right)^\dagger U_t^{H_{BC}}}\\
        &=\opnorm{ U_t^{H_{ABC}}-U_t^{H_{AB}}\left(U_t^{H_{B}}\right)^\dagger U_t^{H_{BC}}}\;.
    \end{split}
    \end{equation}
    We thus have
    \begin{equation}
        \opnorm{ U_t^{H_{ABC}}-U_t^{H_{AB}}\left(U_t^{H_{B}}\right)^\dagger U_t^{H_{BC}}}\leq |B||t|f^\delta(w(B))+|B|^2|t|F_t(w(B)/2)
    \end{equation}
\end{proof}

We are now ready to prove our decomposition theorem ~\ref{Thm:LGA_decomposition}. 
\begin{proof}
    We can apply Lemma \ref{lem:decomposition_Ddim} twice (in 1D, since $C_+^L$ and $C_+^R$ are not connected and $A_+^L$ and $A_+^R$ are not connected, we will need the decomposition four times, but this does not influence our bound). More precisely, we first apply Lemma \ref{lem:decomposition_Ddim} to $U_t^H$ to get an approximate decomposition $U_t^{H_{AB}}(U_t^{C_+})^\dag U_t^{H_{CC_+}}$, and then apply this lemma to $U_t^{H_{AB}}$ to get an approximate decomposition $U_t^{H_B}(U_t^{H_{A_+}})^\dag U_t^{H_{AA_+}}$. To bound the error of the decomposition, note that by definition, $w(C_+)=w(A_+)=w(B)/3$, and we have the trivial bounds $|A_+|<|B|$ and $|C_+|<|B|$, we thus have
    \begin{equation}
        \opnorm{U_t^{H}-U_t^{H_B}\left(U_t^{H_{A_+}+H_{C_+}}\right)^\dagger U_t^{H_{CC_+}+H_{A_+A}}}<2|B||t|f^\delta(w(B)/3)+2|B|^2|t|F_t(w(B)/6)
    \end{equation}
\end{proof}

\section{MI and CMI for commuting-projector models describing topological order }\label{sec:MI&CMI_projector}

In this section, we review the MI and CMI behavior for ground states of commuting-projector models that describe topological order, and show that they vanish for well-separated regions $A$ and $C$.

We start with the MI. Consider a ground state $\rho$ of a commuting projector Hamiltonian $H=\sum_j P_j$. We can construct the projector $\Pi$ to the ground state subspace from $P_j$, \ie $\Pi=\prod_j(1-P_j)$. Topological order is defined by the property of local indistinguishability, \ie $\Pi O\Pi=C_O\Pi$, where $O$ is an operator with support in a contractible region and $C_O$ is a constant that only depends on the operator $O$ \cite{Bravyi_2010,Flammia_2017}.{\footnote{Strictly speaking, this statement only holds for ``fixed-point models" of topological orders. For generic non-fixed-point models of topological orders, this condition only holds up to some errors that vanish in the thermodynamic limit.}} Then for any two observables $O_1$ and $O_2$, such that one of them is supported on a contractible region and no projector $P_j$ can act on their support simultaneously, we can define $\Pi_{1,2}=\prod_{\supp(P_j)\cap\supp(O_{1,2})=\varnothing}(1-P_j)$ and we have
\begin{equation}
\begin{split}
    &\tr(\rho O_1O_2)=\tr(\rho\Pi O_1O_2)=\tr(\rho\Pi O_1O_2\Pi)=\tr(\rho\Pi\Pi_1O_1O_2\Pi_2\Pi)\\
    =&\tr(\rho\Pi O_1\Pi_1\Pi_2O_2\Pi)=\tr(\rho \Pi O_1\Pi O_2\Pi)=\tr(\rho O_1)\tr(\rho O_2)
\end{split}
\end{equation}
where in the last equality, we have used local indistinguishability. This ensures that for any two regions $A$ and $C$ with one of them being contractible, if there is no projector that can act on the two regions simultaneously, we have $\rho_{AC}=\rho_A\otimes\rho_C$ for any ground state $\rho$. Therefore, the mutual information $I_\rho(A:C)=0$.

For the CMI behavior of such a ground state $\rho$, consider the coherent information~\cite{Schumacher_1996, schumacher2001approximatequantumerrorcorrection} and the purification $|\Phi\rangle_{ABCR}$ of $\rho$ where $R$ is the purifying system. Local indistinguishability in the region $A$ guarantees that erasure noise in the region $A$ is correctable. Since the coherent information is preserved under any correctable noise channel, applying this to the erasure of 
$A$ gives:
\begin{equation}
    S_{\Phi}(ABC)-S_{\Phi}(ABCR)=S_{\Phi}(BC)-S_{\Phi}(BCR)\;.
\end{equation}
Using the fact that $|\Phi\rangle$ is a pure state, we thus have
\begin{equation}
    S_{\Phi}(AR)=S_{\Phi}(A)+S_{\Phi}(R)\;.
\end{equation}
Consequently
\begin{equation}
\begin{split}
        I_\rho(A:C|B)&=S_{\rho}(AB)+S_{\rho}(BC)-S_{\rho}(B)-S_{\rho}(ABC)\\
        &=S_{\Phi}(AB)+S_{\Phi}(AR)-S_{\Phi}(B)-S_{\Phi}(R)\\
        &=S_{\Phi}(AB)+S_{\Phi}(A)-S_{\Phi}(B)\\
        &=S_{\rho}(AB)+S_{\rho}(A)-S_{\rho}(B)\;;
\end{split}
\end{equation}
We can now consider a pure ground state $|\psi\rangle$ of the same Hamiltonian $H=\sum_j P_j$. Since $A$ and $AB$ should both be contractible, $\rho$ should be locally indistinguishable with $|\psi\rangle$ on region $AB$, thus
\begin{equation}
\begin{split}
        I_\rho(A:C|B)
        &=S_{\rho}(AB)+S_{\rho}(A)-S_{\rho}(B)=S_{\psi}(AB)+S_{\psi}(A)-S_{\psi}(B)\\
        &=I_{\psi}(A:C)=0\;,
\end{split}
\end{equation}
as long as $\dist(A,C)$ is larger than an $O(1)$ constant.

Therefore, the conditional mutual information $I_\rho(A:C|B)=0$.

\section{MI decay behavior under LGA}\label{sec: MI_decay_pure}

In this section, we present the proof that the mutual information decay is preserved under LGA, \ie the MI part of our Theorem 1.

Suppose $H_0$ is a gapped, almost-local Hamiltonian. By adiabatic continuation, for any state $\rho'$ in the same phase as $H_0$, there exists a state $\rho$ in the ground subspace of $H_0$, such that there exists an almost-local Hamiltonian evolution $U_t^H$ with $\rho'=U_t^H\rho (U_t^H)^\dagger$. The evolution operator $U_t^H$ is determined as in Eqs. \ref{eq:adiabatic1} and \ref{eq:adiabatic2}. By assumption, the MI of $\rho$ decays superpolynomially, \ie for any partition $ABC$ of the lattice as in Fig.~\ref{fig:DimCMI}, 
\begin{equation}
    I_\rho(A:C)=O(\mathrm{poly}(|A|,|B|)\dist(A,C)^{-\infty})\;.
\end{equation}
To study $I_{\rho'}(A:C)$, we can find a reference state $\tilde{\rho}$ close to $\rho'$ with small mutual information. To this end, we decompose $U_t^H$ approximately as in Theorem.~\ref{Thm:LGA_decomposition}, \ie
\begin{equation}
     U_t^{H}\approx \tilde U_t^H\coloneqq U_t^{H_B}\left(U_t^{H_{A_+}+H_{C_+}}\right)^\dagger U_t^{H_{CC_+}+H_{A_+A}}\;.
\end{equation}
Consider the state $\tilde{\rho}=\tilde{U}_t^H\rho(\tilde{U}_t^H)^\dagger$, and note that
\begin{equation}
\tilde{\rho}_{AC}=\tr_{A_+C_+}\left(U_t^{H_{CC_+}+H_{A_+A}}\rho_{AA_+CC_+}(U_t^{H_{CC_+}+H_{A_+A}})^\dagger\right)
\end{equation}
We can bound the MI of $\tilde{\rho}$ using the monotonicity of relative entropy under quantum channels:
\begin{equation}
 \begin{split}
     &I_{\tilde{\rho}}(A:C)=S(\tilde{\rho}_{AC}\|\tilde{\rho}_{A}\otimes\tilde{\rho}_{C})\\
     =&S(\tr_{A_+C_+}\left(U_t^{H_{CC_+}+H_{A_+A}}\rho_{AA_+CC_+}(U_t^{H_{CC_+}+H_{A_+A}})^\dagger\right)\|\tr_{A_+C_+}\left(U_t^{H_{CC_+}+H_{A_+A}}\rho_{AA_+}\otimes\rho_{CC_+}(U_t^{H_{CC_+}+H_{A_+A}})^\dagger\right))\\
     \leqslant & S(\rho_{AA_+CC_+}\|\rho_{AA_+}\otimes\rho_{CC_+})=I_\rho(AA_+:CC_+)
\end{split}   
\end{equation}
We can now utilize the continuity of entropy to derive a bound for MI of $\rho'$. Note that 

\begin{equation}
\sup _{\rho} \frac{1}{2}\left\|\tilde U_t^H\rho(\tilde U_t^H)^\dagger- U_t^H\rho( U_t^H)^\dagger\right\|_1 \leqslant\|\tilde U_t^H-U_t^H\|<\varepsilon(w(B),|B|)
\end{equation}
where $\|A\|_1=\operatorname{Tr} \sqrt{A^{\dagger} A}$ is the 1-norm, $\|\cdot\|$ is the spectral norm. The first inequality can be obtained using the triangle inequality of the trace norm and the H\"older's inequality $\|A B\|_1 \leqslant\|A\|_1\|B\|$ for any $A, B$. By monotonicity of the 1-norm under partial trace, we have $\onenorm{\tilde{\rho}_R-\rho'_R}<2\varepsilon$ for any region $R$.

Note that the mutual information $I_\rho(A:C)=S_\rho(A)+S_\rho(C)-S_\rho(AC)$. By the Fannes–Audenaert inequality, we have
\begin{equation}
    |S_{\tilde{\rho}}(A)-S_{\rho'}(A)|<\varepsilon \log(d^{|A|}-1)+H_2(\varepsilon)
\end{equation}
where we assume the local Hilbert space is $d$-dimensional and $H_2(x)=-x\log x-(1-x)\log(1-x)$. Using the extension of the Fannes inequality for the conditional entropy~\cite{Alicki_2004, Winter_2016}, we also have
\begin{equation}
   \left|\bigl(S_{\tilde{\rho}}(AC)-S_{\tilde{\rho}}(C)\bigr)-\bigl(S_{\rho'}(AC)-S_{\rho'}(C)\bigr)\right|<2\varepsilon \log(d^{|A|})+2\varepsilon H_2(\varepsilon)\;.
\end{equation}
By the triangular inequality, we then obtain
\begin{equation}
    I_{\rho'}(A:C)\leqslant|I_{\rho'}(A:C)-I_{\tilde{\rho}}(A:C)|+I_{\tilde{\rho}}(A:C)<3\varepsilon\log d \cdot|A|+3\varepsilon H_2(\varepsilon)+I_{\rho}(AA_+:CC_+)
\end{equation}

Note that 
\begin{equation}
    \epsilon(w(B),|B|)\coloneqq 2|B||t|f^\delta(w(B)/3)+2|B|^2|t|F_t(w(B)/6)=O(\mathrm{poly}(|B|)\cdot|w(B)|^{-\infty})\;,
\end{equation}
and 
\begin{equation}
    I_{\rho}(AA_+:CC_+)=O(\mathrm{poly}(|AA_+|,|B\backslash A_+C_+|)|\dist(AA_+,CC_+)^{-\infty}|)=O(\mathrm{poly}(|A|,|B|)|\dist(A,C)^{-\infty}|)\;.
\end{equation}
we have thus proved that
\begin{equation}
    I_{\rho'}(A:C)=O(\mathrm{poly}(|A|,|B|)\cdot|w(B)|^{-\infty})
\end{equation}

\section{CMI decay behavior under LGA}\label{sec: CMI_decay_pure}

In this section, we present the proof that the conditional mutual information decay is preserved under LGA, \ie the CMI part of our Theorem.~1.

Suppose $H_0$ is a gapped, almost-local Hamiltonian. By adiabatic continuation, for any state $\rho'$ in the same phase as $H_0$, there exists a state $\rho$ in the ground subspace of $H_0$, such that there exists an almost-local Hamiltonian evolution $U_t^H$ with $\rho'=U_t^H\rho (U_t^H)^\dagger$. By assumption, the CMI of $\rho$ decays superpolynomially, \ie for any partition $ABC$ of the lattice as in Fig.~\ref{fig:DimCMI}, 
\begin{equation}
    I_\rho(A:C|B)=O(\mathrm{poly}(|A|,|B|)\dist(A,C)^{-\infty})\;.
\end{equation}

To study $I_{\rho'}(A:C| B)$, note that a small CMI is equivalent to the existence of an approximate recovery map localized in region $B$ for the erasure noise in region $C$, which we now construct. The main idea is to evolve $\rho'$ back to $\rho$, recover the erasure noise for $\rho$, and evolve it back.

To obtain a clear lightcone structure, we will approximate $U_t^{H}$ with $\tilde U_t^H$ as in Theorem.~\ref{Thm:LGA_decomposition}. Note that
\begin{equation}
\tr_{CC_+}\left((\tilde{U}_t^H)^\dagger(\rho'_{AB}\otimes\pi_C)\tilde{U}_t^H\right)
=\tr_{CC_+}\left((\tilde{U}_t^H)^\dagger \rho'\tilde{U}_t^H\right)
\approx\tr_{CC_+}\left((U_t^H)^\dagger \rho'U_t^H\right)=\rho_{AB\backslash C_+}\;,
\end{equation}
where $\pi_R=\mathbbm{1}/d^{|R|}$ denotes the maximally mixed state on region $R$, and the 1-norm approximation error being $2\epsilon$ from the H\"older's inequality. For $\rho_{AB\backslash C_+}$, the small CMI of $\rho$ ensures that there will be a Petz recovery map supported on $B_-\coloneqq B\backslash (A_+C_+)$, that approximately corrects erasure noise on $CC_+$~\cite{Fawzi-Renner-ACMI}, \ie 
\begin{equation}
    F(\rho, \mathcal{E}_{B_-}^P(\rho_{AB\backslash C_+}))\geqslant 2^{-I_\rho(AA_+:CC_+|B\backslash A_+C_+)/2}\;.
\end{equation}
where the fidelity is defined as $F(\rho,\sigma)\coloneqq\onenorm{\sqrt{\rho}\sqrt{\sigma}}$.
By the Fuchs–van de Graaf inequality, the fidelity can be 2-way bounded by the trace distance, \ie
\begin{equation}\label{eq:fuchs}
    1-F(\rho,\sigma)\leqslant \frac{1}{2}\onenorm{\rho-\sigma}\leqslant \sqrt{1-F^2(\rho,\sigma)}
\end{equation}
we have
\begin{equation}
    \onenorm{\rho-\mathcal{E}_{B_-}^P(\rho_{AB\backslash C_+})}\leqslant 2\sqrt{1-F^2(\rho, \mathcal{E}_{B_-}^P(\rho_{AB\backslash C_+}))}\leqslant 2\sqrt{\ln 2 \, I_\rho(AA_+:CC_+|B\backslash A_+C_+) }
\end{equation}
where we have used that $1-2^{-x}\leqslant x\ln 2$ for $x\geqslant 0$ in the last step.

We can then evolve $\rho$ back to $\rho'$ to finish the recovery. Note that
\begin{equation}
    \onenorm{\rho'-\tilde{U}_t^H\rho(\tilde{U}_t^H)^\dagger }<2\varepsilon(w(B),|B|)\;.
\end{equation}
By the monotonicity of one norm under quantum channels and the triangular inequality, we thus have
\begin{equation}
\begin{split}
    &\onenorm{\rho'-\tilde{U}_t^H\left(\mathcal{E}_{B_-}^P\circ\tr_{CC_+}\left((\tilde{U}_t^H)^\dagger(\rho'_{AB}\otimes\pi_C)\tilde{U}_t^H\right)\right)(\tilde{U}_t^H)^\dagger}\\
    \leqslant &\onenorm{\rho'-\tilde{U}_t^H\rho(\tilde{U}_t^H)^\dagger }+\onenorm{\rho-\mathcal{E}_{B_-}^P(\rho_{AB\backslash C_+})}+\onenorm{\rho_{AB\backslash C_+}-\tr_{CC_+}\left((\tilde{U}_t^H)^\dagger(\rho'_{AB}\otimes\pi_C)\tilde{U}_t^H\right)}\\
    < & 4\varepsilon(w(B),|B|)+2\sqrt{\ln 2 \, I_\rho(AA_+:CC_+|B\backslash A_+C_+) }\\
    =&O(\mathrm{poly}(|A|,|B|)\cdot|w(B)|^{-\infty})
\end{split}
\end{equation}

Note that the structure of $\tilde{U}_t^H$ enables cancellations such that the recovery channel can be localized on region $B$, \ie
\begin{equation}
    \tilde{U}_t^H\left(\mathcal{E}_{B_-}^P\circ\tr_{CC_+}\left((\tilde{U}_t^H)^\dagger(\rho'_{AB}\otimes\pi_C)\tilde{U}_t^H\right)\right)(\tilde{U}_t^H)^\dagger=
\tilde{\mathcal{E}}^P_B(\rho'_{AB})\;,
\end{equation}
where
\begin{equation}
    \tilde{\mathcal{E}}^P_B\coloneqq \Ad_{U^{H_B}_t(U^{H_{C_+}}_t)^\dagger U^{H_{CC_+}}_t}\circ \mathcal{E}^P_{B_-}\circ \tr_{C_+}\circ\Ad_{(U^{H_{B}}_t)^\dagger}
\end{equation}
with $\Ad_U(\cdot)\coloneqq U\cdot U^\dagger$.

The existence of a good recovery channel provides an upper bound for the CMI~\cite{Fawzi-Renner-ACMI}, \ie
\begin{equation}
    I_{\rho'}(A:C|B)\leqslant 7 \log d \cdot |A|\sqrt{\onenorm{\rho'-\tilde{\mathcal{E}}^P_B(\rho'_{AB})}/2}=O(\mathrm{poly}(|A|,|B|)\cdot|w(B)|^{-\infty})
\end{equation}
we thus proved that
\begin{equation}
    I_{\rho'}(A:C|B)=O(\mathrm{poly}(|A|,|B|)\cdot|w(B)|^{-\infty})\;.
\end{equation}

\section{MI and CMI decay for mixed-state phases}\label{sec: CMI_decay_mixed}

In this section, we generalize our results from closed quantum systems to mixed-state phases in open systems. For convenience, we review our definition of mixed-state phases based on locally reversible finite-depth channels. 

\begin{definition}[Definition 1 in the main text]\label{def:mixed_reversibility}
      Two states $\rho$ and $\rho'$ are in the same phase if there exist local channel circuits $\mathcal{C}=\mathcal{C}_T \cdots \mathcal{C}_2 \mathcal{C}_1$ and $\tilde{\mathcal{C}}=\tilde{\mathcal{C}}_1\tilde{\mathcal{C}}_2\cdots \tilde{\mathcal{C}}_T $  (each $\mathcal{C}_t$ or $\tilde{\mathcal{C}}_t$ is a layer of non-overlapping local channel gates) such that:
     \begin{equation} \label{eq: two-way connection}
         \mathcal{C}(\rho)=\rho',
         \quad
         \tilde{\mathcal{C}}(\rho')=\rho.
     \end{equation}
    We also require the channels to be locally reversible, \ie for any $t$ and any $\mathcal{C}^R_t$ and $\tilde{\mathcal{C}}^R_t$ being a layer composed of a subset of gates in $\mathcal{C}_t$ and $\tilde{\mathcal{C}}_t$, respectively, with the supports of the gates fully contained in a region $R$:
    \begin{align} \label{eq: local reversibility 1}
    \tilde{\mathcal{C}}^R_t\mathcal{C}^R_t\left(\mathcal{C}_{t-1} \cdots \mathcal{C}_2 \mathcal{C}_1(\rho)\right)=\mathcal{C}_{t-1} \cdots \mathcal{C}_2 \mathcal{C}_1(\rho)\;,\\
    \label{eq: local reversibility 2}
    \mathcal{C}^R_t\tilde{\mathcal{C}}^R_t\left(\tilde{\mathcal{C}}_{t-1} \cdots \tilde{\mathcal{C}}_2 \tilde{\mathcal{C}}_1(\rho')\right)=\tilde{\mathcal{C}}_{t-1} \cdots \tilde{\mathcal{C}}_2 \tilde{\mathcal{C}}_1(\rho')\;.
    \end{align}
\end{definition}


We claim that the decay behavior of CMI and MI is preserved for states in the same phases, as per Definition \ref{def:mixed_reversibility}, \ie
\begin{theorem}[Theorem.~2 in the main text]\label{thm:mixed_CMI_decay}
Let $\rho$ and $\rho'$ be two mixed states belonging to the same phase. If for any partition $ABC$ of the lattice with $A$ being a contractible region shielded from $C$ by the region $B$, either of the following two conditions is satisfied, 
\begin{align}\label{eq:decay2}
    I_\rho(A:C)=O(\mathrm{poly}(|A|,|B|)\dist(A,C)^{-\infty})\;;\\
    \label{eq:decay1}
    I_\rho(A:C| B)=O(\mathrm{poly}(|A|,|B|)\dist(A,C)^{-\infty})\;,
\end{align}
then the same condition holds for $\rho'$.

This statement still holds if the superpolynomial decay behaviors in $\dist(A, C)$ are replaced by polynomial or exponential decay behaviors.
\end{theorem}

\begin{figure}
    \centering
    \includegraphics[width=1\linewidth]{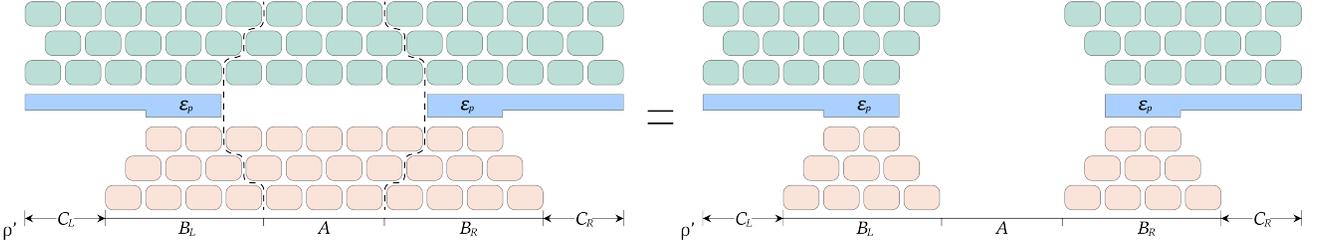}
\caption{Construction of the recovery map for $\rho’$. (a) We consider the pullback of the Petz recovery map $\mathcal{E}_P$. Note that while the output of $\mathcal{E}_P$ is on $BC\backslash A_+$, $\supp(\mathcal{E}_P)\subseteq B\backslash (A_+C_+)$. We first evolve $\rho'_{AB}$ with $\mathcal{C}$, obtaining $\rho_{AB\backslash C_+}$ when erasing $C_+$, and apply the Petz recovery map to get $\rho$. The original $\rho'$ is obtained by evolving $\rho$ back with $\tilde{\mathcal{C}}$. (b) The gates outside of the lightcone of $\mathcal{E}_P$ can be canceled out by the local reversibility condition, thus the recovery map we constructed on the left hand side can be replaced by a local recovery map $\mathcal{E}'$ as in the right hand side. }
    \label{fig:recovery}
\end{figure}

\begin{proof}

    We first consider the MI, where Eq.~(\ref{eq: two-way connection}) is enough to prove the Theorem.~\ref{thm:mixed_CMI_decay} and Eqs.~(\ref{eq: local reversibility 1}) and~(\ref{eq: local reversibility 2})  are not necessary. Suppose for $\rho$, $I_{\rho}(AA_+:CC_+)$ has a certain decay behavior, then for $\rho'=\mathcal{C}(\rho)$ note that
\begin{equation}
\begin{split}
    \rho'_{AC}&=\tr_B(\mathcal{C}(\rho_{AA_+CC_+}\otimes\pi_{B\backslash (A_+C_+)}))\\
    \rho'_{A}\otimes\rho'_{C}&=\tr_B(\mathcal{C}(\rho_{AA_+}\otimes\rho_{CC_+}\otimes\pi_{B\backslash (A_+C_+)}))\;.
\end{split}
\end{equation}
We can bound the MI of $\rho'$ using the monotonicity of relative entropy under quantum channels:
\begin{equation}
\begin{split}
    &I_{\rho'}(A:C)=S(\rho'_{AC}\|\rho'_A\otimes\rho_C')\\
    =&S(\tr_B(\mathcal{C}(\rho_{AA_+CC_+}\otimes\pi_{B\backslash (A_+C_+)}))\|\tr_B(\mathcal{C}(\rho_{AA_+}\otimes\rho_{CC_+}\otimes\pi_{B\backslash (A_+C_+)})))\\
    \leqslant&S(\rho_{AA_+CC_+}\otimes\pi_{B\backslash (A_+C_+)})\|\rho_{AA_+}\otimes\rho_{CC_+}\otimes\pi_{B\backslash (A_+C_+)}))\\
    =&S(\rho_{AA_+CC_+}\|\rho_{AA_+}\otimes\rho_{CC_+})\\
    =&I_\rho(AA_+:CC_+)
\end{split}
\end{equation}
where the relative entropy $S(\rho\|\sigma)=\tr(\rho\log \rho-\rho\log \sigma)$.

We thus obtained if 
\begin{equation}
    I_{\rho}(AA_+:CC_+)=O(\mathrm{poly}(|AA_+|,|B\backslash A_+C_+|)|\dist(AA_+,CC_+)^{-\infty}|)\;, 
\end{equation}
we are guaranteed that
\begin{equation}
    I_{\rho'}(A:C)=O(\mathrm{poly}(|AA_+|,|B\backslash A_+C_+|)|\dist(AA_+,CC_+)^{-\infty}|)=O(\mathrm{poly}(|A|,|B|)|\dist(A,C)^{-\infty}|)\;.
\end{equation}
This still holds if the superpolynomial decay is replaced by polynomial or exponential decay.
\\

    Now we turn to the CMI. To study $I_{\rho'}(A:C|B)$ with the partitions of regions $ABC$ as in Fig.~\ref{fig:DimCMI}, we consider $AA_+^{(t)}$ to be the lightcone of region $A$ and $CC_+^{(t)}$ to be the lightcone of region $C$, under a depth-$t$ local channel. More explicitly, if the local channels are $k$-local, we can take:
    \begin{align}
        A_+^{(t)}&=\{j\in B|\dist(j,A)\leqslant(k-1)(t-1)\}\\
        C_+^{(t)}&=\{j\in B|\dist(j,C)\leqslant(k-1)(t-1)\}\\
        B_-^{(t)}&=B\backslash A_+^{(t)}C_+^{(t)}\;.
    \end{align}
    Note that when $t=T$, $A_+^{(T)}=A_+$, $C_+^{(T)}=C_+$ and $B_-^{(T)}=B_-$ are reduced to the notation of $A_+$, $C_+$ and $B_-$ we defined in the main text. 
    
    Suppose $AA_+$ and $CC_+$ are well-separated, then for $\rho$, 
    \begin{equation}
    I_{\rho}(AA_+:CC_+|B\backslash A_+C_+)=O(\mathrm{poly}(|AA_+|,|B\backslash A_+C_+|)|\dist(AA_+,CC_+)^{-\infty}|)\;, 
    \end{equation}
    For notational simplicity, we can denote the left-hand side by $I$. Thus the for erasure error on region $CC_+$ for state $\rho$, one can construct a Petz recovery map $\mathcal{E}^P$ supported on $B\backslash A_+C_+$, such that~\cite{Fawzi-Renner-ACMI}
    \begin{equation}\label{eq_supp:petzVsCMI}
    F(\rho,\mathcal{E}^P(\tr_{CC^+}\rho))\geqslant 2^{-I/2}\;.
    \end{equation}
    Such a recovery map can induce a recovery map for $\rho'$ with erasure noise in the region $C$. To see this, we apply the same idea as for the gapped ground states. Note that
    \begin{equation}
        \tr_{CC_+}(\mathcal{C}(\rho'_{AB}\otimes\pi_C))=\tr_{CC_+}\mathcal{C}(\rho')=\rho_{AB\backslash C_+}
    \end{equation}
    where $\pi_C=\mathbbm{1}/d^{|C|}$ denotes the maximally mixed state in region $C$. We can then reconstruct the state $\rho$ by the Petz recovery map and evolve it back to $\rho'$, with the fidelity estimated by (\ref{eq_supp:petzVsCMI}), \ie
    \begin{equation}\label{eq_supp:RecoveryPullBack}
        F(\rho',\tilde{\mathcal{C}}\circ \mathcal{E}_P \circ\tr_{CC_+}(\mathcal{C}(\rho'_{AB}\otimes\pi_C)))\geqslant 2^{-I/2}
    \end{equation}
    where we have used the fact that the fidelity is monotonic for channels. 
    
    While the recovery channel in (\ref{eq_supp:RecoveryPullBack}) is global, it can be reduced to a local recovery via the local reversibility condition.  More explicitly, by the lightcone property, 
    \begin{equation}
        \tr_{CC_+}(\mathcal{C}(\rho'_{AB}\otimes\pi_C))=\tr_{C_+}\mathcal{C}_T^{\overline{C_+^{(T)}}}\circ\mathcal{C}_{T-1}^{\overline{C_+^{(T-1)}}}\circ\cdots \circ \mathcal{C}_1^{\overline{C_+^{(1)}}}(\rho_{AB}')
    \end{equation}
    We thus have 
\begin{equation}
\begin{split}
    &\tilde{\mathcal{C}}\circ \mathcal{E}_P\circ \tr_{CC_+}(\mathcal{C}(\rho'_{AB}\otimes\pi_C))\\
    =&\tilde{\mathcal{C}}_1\circ\cdots\circ \tilde{\mathcal{C}}_T^{\overline{A_+^{(T)}}}\circ\tilde{\mathcal{C}}_T^{A_+^{(T)}}\circ\mathcal{E}_P\circ\tr_{C_+}\mathcal{C}_T^{\overline{C_+^{(T)}}}\circ\mathcal{C}_{T-1}^{\overline{C_+^{(T-1)}}}\circ\cdots \circ \mathcal{C}_1^{\overline{C_+^{(1)}}}(\rho_{AB}')\\
    =&\tilde{\mathcal{C}}_1\circ\cdots\circ \tilde{\mathcal{C}}_T^{\overline{A_+^{(T)}}}\circ\mathcal{E}_P\circ\tr_{C_+}\mathcal{C}_{T}^{B_-^{(T)}}\circ\mathcal{C}_T^{\overline{C_+^{(T-1)}}}\circ\cdots \circ \mathcal{C}_1^{\overline{C_+^{(1)}}}(\rho_{AB}')\\
    =&\cdots=\mathcal{E}'(\rho_{AB}')\coloneqq\tilde{\mathcal{C}}_1^{\overline{A_+^{(1)}}}\circ\cdots\circ \tilde{\mathcal{C}}_T^{\overline{A_+^{(T)}}}\circ\mathcal{E}_P\circ\tr_{C_+}\mathcal{C}_{T}^{B_-^{(T)}}\mathcal{C}_{T-1}^{B_-^{(T-1)}}\circ\cdots \circ \mathcal{C}_{1}^{B_-^{(1)}}(\rho_{AB}')\;,
\end{split}
\end{equation}
where in the third line we use the local reversibility condition to cancel the gates in one layer of the circuit, and in the fourth line we repeat this procedure to get a local recovery channel (see Fig.~\ref{fig:recovery} as an illustration of this construction). Notice that $\mathcal{E}'$ is supported on $B\backslash A_+^{(1)}C_+^{(1)}=B$, so there exist a recovery map of erasure noise on $C$ for state $\rho'$ with fidelity $2^{-I/2}$.

The existence of a good recovery channel provides an upper bound for the CMI~\cite{Fawzi-Renner-ACMI}, \ie 
\begin{equation}
\begin{split}
    I_{\rho'}(A:C|B)&\leqslant 7 \log d \cdot|A|\sqrt{\onenorm{\rho'-\mathcal{E}'(\tr_{C}\rho')}/2}\\
    &\leqslant 7 \log d\cdot |A| \left(1-F^2(\rho',\mathcal{E}'(\tr_{C}\rho'))\right)^{1/4}\\
    &\leqslant 7(\ln2)^{1/4} \log d\cdot |A| I^{1/4}\;,
\end{split}
\end{equation}
where in the third line we have used the fact that $1-2^{-x}\leqslant x\ln2$ for $x\geqslant 0$.

We thus obtained if 
\begin{equation}
    I=I_{\rho}(AA_+:CC_+|B\backslash A_+C_+)=O(\mathrm{poly}(|AA_+|,|B\backslash A_+C_+|)|\dist(AA_+,CC_+)^{-\infty}|)\;, 
\end{equation}
we are guaranteed that
\begin{equation}
    I_{\rho'}(A:C|B)=O(\mathrm{poly}(|AA_+|,|B\backslash A_+C_+|)|\dist(AA_+,CC_+)^{-\infty}|)=O(\mathrm{poly}(|A|,|B|)|\dist(A,C)^{-\infty}|)\;.
\end{equation}
Thus a locally reversible finite depth quantum channel preserves a superpolynomially decaying CMI. This still holds if the superpolynomial decay is replaced by polynomial or exponential decay.

\end{proof}

\section{Entanglement inequalities for fermionic systems} \label{secapp: fermion}

For completeness, in this section we show that the entanglement inequalities we used in previous sections also apply to the fermionic systems. For our results to hold, we only need to check the Fannes-type inequalities and the existence of an approximate Petz recovery map for small enough CMI.

For a tripartite fermionic state $\rho$ defined on regions $A,B,C$, we consider an ordering of the lattice sites $j$ such that for any $j_{A,B,C}\in A,B,C$, we have $j_A<j_B<j_C$. We consider the Jordan-Wigner transformation that corresponds to this specific ordering, \ie
\begin{equation}
    a_j=\frac{1}{2}(X_j+iY_j)\prod_{j'<j}Z_{j'}\;.
\end{equation}
where $a_j$ is the annihilation operator for the fermion at the site $j$.

We now consider the reduced density matrix $\rho_R$ on a local region $R$ in terms of this ordering of $j$, \ie $R=A, B,C, AB, BC, ABC$. The super-selection rule requires $\rho_R$ to be parity even, \ie $[\rho_R,\Pi_R]=0$, where $\Pi_R=(-1)^{\sum_j \hat{n}_j}$ is the parity operator on region $R$. We can thus expand $\rho_R$ in terms of even number of $a_j$ and $a^\dagger_j$ in $R$, thus after the Jordan-Wigner transformation $\rho_R\rightarrow \rho_R'$, we still have a local density matrix (of spin system) on region $R$. If one writes $\rho_R$ in the mode occupation basis and $\rho_R'$ in computational basis, their matrix form will be the same. Thus for any two tripartite fermionic states $\rho$ and $\sigma$, we have
\begin{equation}
    S(\rho_R)=S(\rho_R'),\; \quad  S(\sigma_R)=S(\sigma_R'),
\end{equation}
\begin{equation}
    \onenorm{\rho_R-\sigma_R}=\onenorm{\rho'_R-\sigma'_R},
\end{equation}
\begin{equation}
    F(\rho_R,\sigma_R)=F(\rho'_R,\sigma'_R).
\end{equation}
Thus the Fannes-type inequalities still hold for fermionic systems.

We now turn to the existence of an approximate Petz recovery map. Note that the erasure noise on $A$ for $\rho_{ABC}$ is equivalent to the erasure noise on $A$ for $\rho'_{ABC}$, and $I_\rho(A:C|B)=I_{\rho'}(A:C|B)$. A small CMI for $\rho'$ thus ensures the existence of a rotated Petz recovery map~\cite{Junge_2018, Wilde_2015,Hu_2024} for $\rho'_{ABC}$, \ie
\begin{equation}
    \mathcal{D}_{t}\left(\rho'_{BC}\right)=\rho_{AB}'^{\frac{1+i t}{2}} \rho_B'^{\frac{-1-i t}{2}} \rho'_{BC} \rho_B'^{\frac{-1+i t}{2}} \rho_{AB}'^{\frac{1-i t}{2}}
\end{equation}
with
\begin{equation}
    \max_t F(\mathcal{D}_{t}\left(\rho'_{BC}\right),\rho'_{ABC})\geqslant 2^{-I_{\rho'}(A:C|B)/2}\;.
\end{equation}
Note that $\mathcal{D}_t$ is written purely in terms of the density matrices supported in their respective regions, so under the inverse Jordan-Wigner transformation $\mathcal{D}_t$ remains a local channel with input supported in $B$ and output supported on $AB$. Therefore, a small CMI for $\rho$ also guarantees the existence of an approximate Petz recovery map for the original fermionic system.

It is also worth noting that here we do not use the standard approximate Petz recovery map since it is not uniquely determined by the reduced density matrices of $\rho'$ on different regions, so under the inverse Jordan-Wigner transformation this map is not guaranteed to preserve its locality. On the other hand, the rotated Petz recovery map preserves the locality under the inverse Jordan-Wigner transformation.

\bibliography{lib.bib}